\documentclass[10pt,journal,compsoc,cspaper,twoside]{IEEEtran}

\usepackage[utf8]{inputenc}
\usepackage{multicol}
\usepackage{array}
\usepackage{hhline}
\usepackage{amsmath}
\usepackage{adjustbox}
\usepackage{wrapfig}
\usepackage{booktabs} 
\usepackage{graphicx}
\usepackage{xspace}
\usepackage{mathtools}
\usepackage{soul}
\usepackage{multirow}
\usepackage[caption=false,font=normalsize,labelfont=sf,textfont=sf]{subfig}
\usepackage{algorithmicx}
\usepackage{comment}
\usepackage{etoolbox}
\usepackage{xcolor}
\usepackage{eucal}
\usepackage{balance}
\usepackage{threeparttable}
\usepackage{wasysym}
\usepackage{MnSymbol} 
\usepackage{orcidlink}
\usepackage{tikz}
\usepackage{enumitem}
\usetikzlibrary{shapes.geometric, arrows}

\usepackage{flushend} 

\hypersetup{
    colorlinks,
    citecolor=black,
    filecolor=black,
    linkcolor=black,
    urlcolor=black
}

\newcommand{\shortname}{\textsc{H2Auth}\xspace}

\newcommand{\audiorelay}{audio relay\xspace}

\newcommand{\eg}{e.\,g.,\ }
\newcommand{\ie}{i.\,e.,\ }

\newcommand{\aaf}{\vspace*{-6pt}}
\newcommand{\af}{\vspace*{-3pt}}

\DeclareMathAlphabet{\mathpzc}{OT1}{pzc}{m}{it}

\usepackage{pifont}

\newcolumntype{P}[1]{>{\centering\arraybackslash}p{#1}}
\newcommand{\para}[1]{\vspace{3pt}\noindent\textbf{#1}}
\newcommand{\Para}[1]{\noindent\textbf{#1}}

\usepackage{url}
\makeatletter
\g@addto@macro{\UrlBreaks}{\UrlOrds}
\makeatother

\def\BibTeX{{\rm B\kern-.05em{\sc i\kern-.025em b}\kern-.08em
    T\kern-.1667em\lower.7ex\hbox{E}\kern-.125emX}}

\usepackage{balance}

\begin{document}

\title{Turning Noises to Fingerprint-Free ``Credentials'':\\Secure and Usable Drone Authentication}

\author{Chuxiong Wu\,\orcidlink{0000-0003-0243-661X}, \IEEEmembership{Student Member, IEEE}, and 
Qiang Zeng\,\orcidlink{0000-0001-9432-6017}
 \IEEEcompsocitemizethanks{
    \IEEEcompsocthanksitem 
    Both authors are with the Department of Computer Science, George Mason University, Fairfax, VA, 22032. E-mail: \{cwu27, zeng\}@gmu.edu
 }
}

\markboth{IEEE TRANSACTIONS ON MOBILE COMPUTING,~VOL.~X,~NO.~Y,~MONTH~YEAR}{Wu et al.: Turning Noises to Fingerprint-Free ``Credentials'': Secure and Usable Drone Authentication}

\IEEEtitleabstractindextext{
\begin{abstract}
Drones have been widely used in various services, such as delivery and surveillance. 
Authentication forms the foundation of the security of these services.
However, drones are expensive and may carry important payloads.
To avoid being captured by attackers, drones should keep a safe distance from the verifier before authentication succeeds. This makes
authentication methods that only work in very close proximity not applicable.
Our work leverages drone noises for authentication.
While using sounds  for authentication is highly usable,
how to handle various attacks that manipulate sounds is an \emph{unresolved challenge}.
It is also unclear how to ensure robustness under various environmental sounds.
Being the first in the literature, we address the two major challenges by exploiting unique characteristics of drone noises.
We thereby build an authentication system 
that does \emph{not} rely on any drone sound fingerprints, keeps resilient to attacks, and is robust under environmental sounds.
An extensive evaluation demonstrates its security and usability. 
\end{abstract}

}

\maketitle

\IEEEraisesectionheading{\section{Introduction}\label{sec:Introduction}}
\IEEEPARstart{B}{ecause} of multiple advantages, such as fast speed, low manual cost, and access to remote areas, drones have been widely used in various services, such as delivery, surveillance, telecommunication, etc~\cite{hassanalian2017classifications}.
However, the growing popularity of drone services renders them an appealing target of attacks~\cite{wazid2018authentication}. 
For instance, malicious drones could mimic legitimate ones to gain unauthorized access to restricted locations and resources. 
Likewise, attackers might impersonate legitimate users to exploit drone services, such as stealing packages carried by a delivery drone~\cite{wu2022g2auth}.

Authentication
plays a fundamental role in ensuring security. For example, a user needs to authenticate a drone before trusting it to pick 
up a package. Likewise, a delivery drone needs to authenticate a user before releasing a package. Moreover, in scenarios where a drone enters a warehouse, collaborates with a robot, acts as an agent for a customer, or approaches a sensitive area,  
authentication serves as the foundation of security~\cite{wu2022g2auth,sharp2022authentication}.

Compared to many authentication scenarios, drone authentication is unique~\cite{wu2022g2auth}. Drones are expensive and may carry important data and/or payloads. To avoid being captured by attackers, drones should keep a safe distance from the verifier before authentication succeeds. This makes authentication methods that work in very close proximity, such as NFC and keypads, not usable.

The motors and propellers of each drone generate unique noises due to manufacturing imperfections~\cite{sounduav}.
Thus, drone noises present an opportunity for authentication. 
Our observation is that if a drone and a verifier are in proximity in the physical world,
both should be able to record the noises of the drone.
We propose that, when a drone hovers,  
the drone and the verifier (e.g., the smartphone of a user) both record the drone noises. After exchanging the recordings, the drone and the verifier can independently
check  the similarity of the two recordings for mutual authentication. 
Our work has the following advantages.
(1) \textbf{Zero drone-sound fingerprints}. Sound features of a drone vary when the weight of its payload changes~\cite{ibrahim2022noise2weight}, affecting the sound fingerprints. Unlike prior work~\cite{sounduav, daaf}, our approach does not rely on any sound fingerprints. 
(2) \textbf{Conducting mutual authentication}. As the drone and the verifier independently check the similarity
of the exchanged recordings, mutual authentication can be attained. 
(3) \textbf{Resilience to attacks.} An attacker can record and replay a drone's sounds to fool approaches that relies on drone noises for authentication. 
Such attacks are not examined in prior work, but carefully studied in our work.
(4) \textbf{Robustness to environmental sounds.} 
How to ensure the sound-based authentication to be robust to various environmental noises is not examined in prior work but studied in this work. 

Our work addresses the following two major challenges.
The first challenge is to handle
various attacks against sounds.
(1) \textbf{Dominant sound attack:} an attacker can impose 
the same \emph{dominant} sounds  near both the verifier and
the drone
to fool the authentication. (2) \textbf{Audio relay attack:}
the sounds near the verifier (or drone, resp.) can be recorded and replayed near the drone (or verifier, resp.).

The other challenge is that drone services are conducted under a variety of environmental sounds, such as traffic noises, people chatting, radio news, 
and music. 
Moreover, the microphone on the drone side
mainly records the drone noises due to the close proximity to the motors and propellers, 
while what is recorded by the verifier tends to be affected by environmental sounds.
The significant disparity is a barrier to accurate sound comparison.

To address these challenges,
our insight is that
a secure communication channel can be established using mature techniques, such as PKI. Instead of collecting a drone's noise fingerprint~\cite{sounduav, daaf}, the drone and the verifier use
the secure channel to exchange the recordings of the \emph{current} sounds. 
We devise a novel method 
for accurate and robust audio comparison, which tackles
the disparity aforementioned.  Plus, effective countermeasures that exploit
the uniqueness of drone noises are developed to defeat attacks. 
        
We build a system, named Hum2Auth (\shortname), and perform an extensive evaluation. Below is a subset of the questions the evaluation studies. \emph{Is the accuracy high?} \emph{Can the system be used for different drones and verifiers}? \emph{Is it resilient to attacks?} \emph{Can it work under various environmental sounds?} 
The evaluation results give positive answers to all the questions. 
We make the following contributions.
\begin{itemize}
    \item We propose a highly secure and usable mutual authentication solution for drone services that does \textbf{not} rely on any drone noise fingerprints.

    \item 
    This is the first work that defeats 
     various attacks against a sound-based authentication system and keeps resilient to various environmental sounds, which
     distinguishes our work from prior work.    We devise a novel audio similarity comparison method that exploits the characteristics of drone noises to cope with the significant sound disparity between drones and verifiers, attaining a high accuracy (=0.997). 
    
  \item This is the first mutual authentication approach for drones without relying on biometrics. A robot, car, smart doorbell or garage can work as the verifier,
  while prior state of the art work requires a human being to be the present.
    
    \item We implement a prototype and conduct an extensive
    evaluation, demonstrating its security and usability. 
    
\end{itemize}
\section{Design Choices and Threat Model} \label{Sec:design_threat}

\subsection{Design Choices} \label{sec:background}
We first discuss some straightforward design choices.

\begin{figure}
\centering
\hfill
\subfloat[Car]{\includegraphics[scale=0.30]{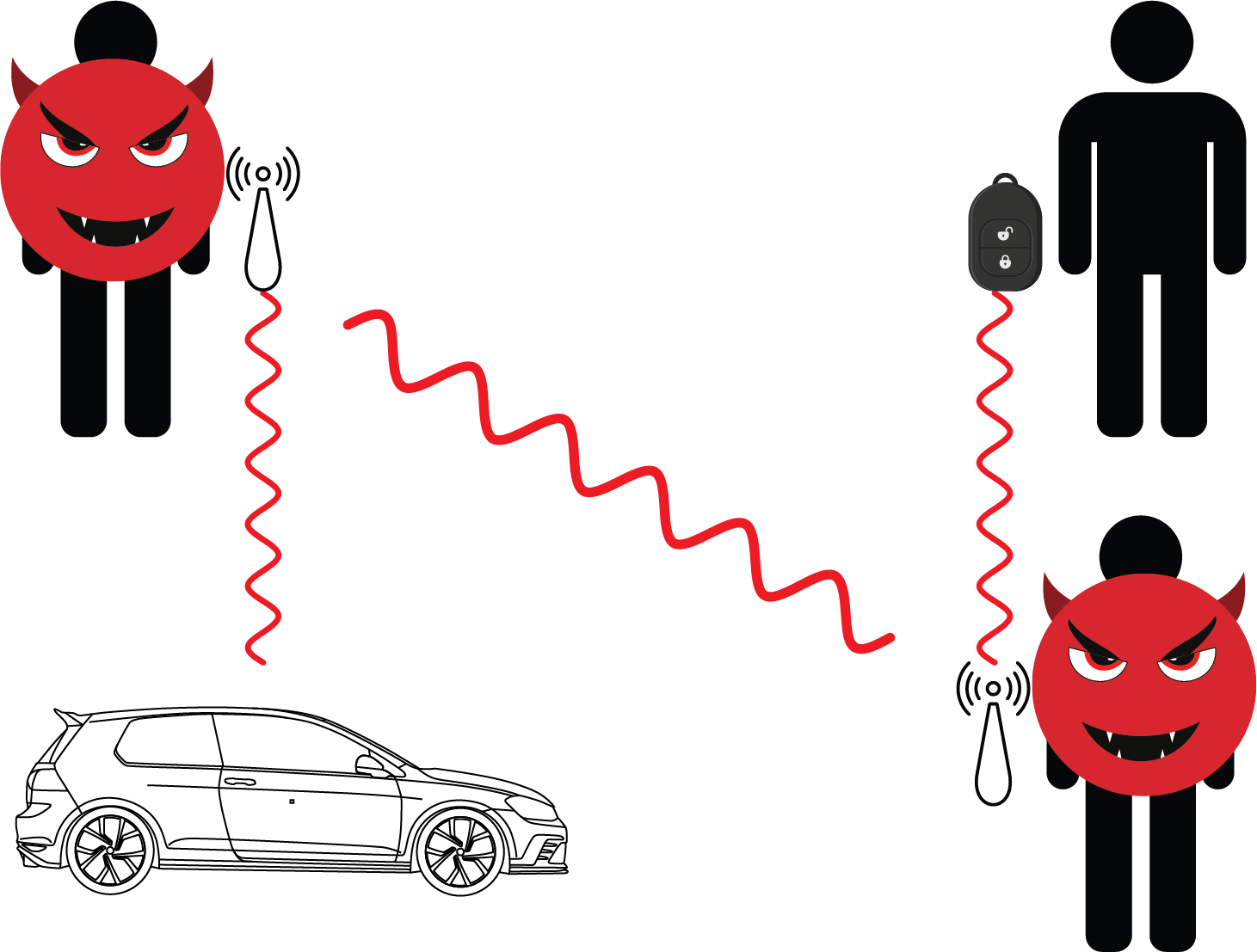}} 
\hfill
\subfloat[Drone]{
\includegraphics[scale=0.30]{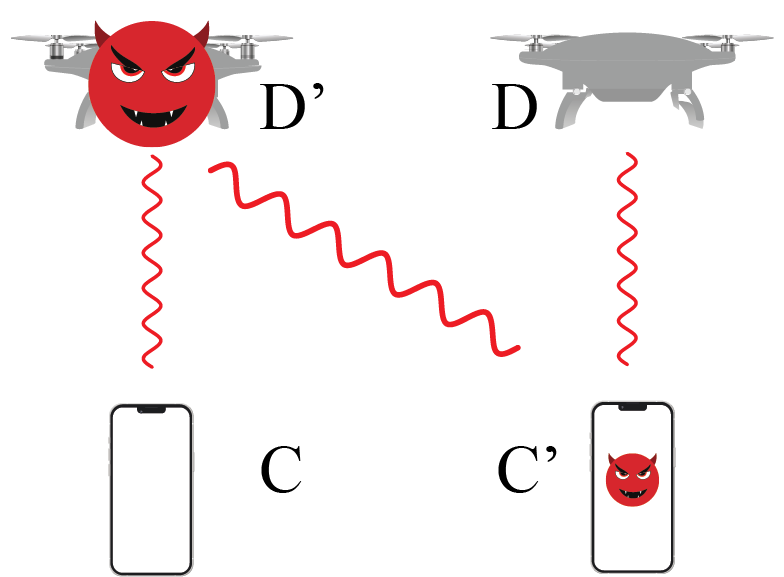}}
\caption{Radio relay attacks.}
\label{fig:relay_attacks}
\end{figure}

\para{Bluetooth.} Much research has demonstrated the insecurity of intuitively using radios for verifying proximity. 
For instance, in the case of reduced-range Bluetooth,  even if the communication channel is protected by a key, attackers can launch \textbf{radio relay attacks} (aka \emph{Mafia Fraud Attacks}~\cite{desmedt1987special}) without breaking the underlying cryptography key. 
Such attacks have been used to compromise Passive Keyless Entry and Start (PKES) systems used in modern cars~\cite{francillon2011relay}.
As shown in Figure~\ref{fig:relay_attacks}(a), the attack involves two attackers working together. One attacker stands near the targeted car, while the other stays in proximity to the car owner, equipped with a device capable of relaying the signal from the key fob to the other attacker and further to the car, without the need to crack cryptographic secrets. Car thefts applying relay attacks have been reported~\cite{real} and are
cheap (\$22)~\cite{cheap}.
Readers are referred to \cite{ranganathan2017we,danev2010attacks} about the insecurity of other naive methods for verifying proximity, such as RSSI, radio fingerprinting, etc.

Like attacks against cars~\cite{francillon2011relay}, radio relay attacks against
drone authentication can also be launched~\cite{sharp2022authentication}. 
For example, a family who picnics in Central Park has ordered an expensive bottle of wine; due to GPS navigation inaccuracy~\cite{gps_accuracy} or spoofing~\cite{kerns2014unmanned, shepard2012evaluation,shen2020drift,zeng2018all}, as shown in Figure~\ref{fig:relay_attacks}(b), a drone $D$ that delivers the bottle
hovers near an attacker, who controls a malicious device $C'$.
The attacker, meanwhile, controls a malicious 
drone $D'$ to hover in front of the legitimate user, who mistakenly considers $D'$ as the service drone and starts
the authentication procedure (e.g., sending a purchase code or PIN from $C$~\cite{qualcomm-code}).
Then, $D'$ relays the radio signal, \emph{without knowing the encryption key}, to $C'$, which relays it to $D$.

Compared to attacks in the car scenario, attacks in the drone scenario do not need an attacker to personally approach
and stay close to the victim user. Instead, they send a rogue drone to a user expecting an authentic drone.

\para{QR code.} A Google's patent~\cite{google-qr} has the drone $D$ authenticate a user by scanning a QR code shown on the user's smartphone $C$. However, it is vulnerable to \emph{vision relay attacks}~\cite{wu2022g2auth}:
as shown in Figure~\ref{fig:relay_attacks}(b), the malicious drone $D'$ scans the QR code from $C$ and sends it to the malicious smartphone $C'$, which displays the same QR code to $D$. Readers are referred to \cite{wu2022g2auth} about the insecurity of an enhanced version of QR code, that is, quickly switching multiple
QR code.

\para{Distance bounding.} 
Distance bounding~\cite{brands1993distance} enables one device to securely establish an upper bound on its distance to another device, which can be used to verify proximity for authentication. However, as they are based on the time difference between sending challenge bits and receiving the corresponding response bits, the accuracy is sensitive to the slightest processing latency. Thus, 
it requires special hardware~\cite{rasmussen2010realization}, which is not widely available. Indeed, it is unfair to require low-income people
to purchase high-end smartphones that support distance bounding in order to
benefit from the advances of technologies, such as drone authentication. The security of distance bounding protocols is still being actively studied~\cite{mauw2018distance,cremers2012distance,avoine2018security}. As an interoperable ecosystem for distance bounding is not yet available~\cite{fira}, compatibility and interoperability issues between drones and user-side devices cannot be ignored.

These straightforward designs are either insecure or not usable (e.g., requiring 
special hardware on the user side). We aim at a highly secure and usable authentication
method. 

\begin{figure*}
\captionsetup[subfigure]{justification=centering}
\centering
\subfloat[Dominant sound attack]
{\includegraphics[scale=0.14]{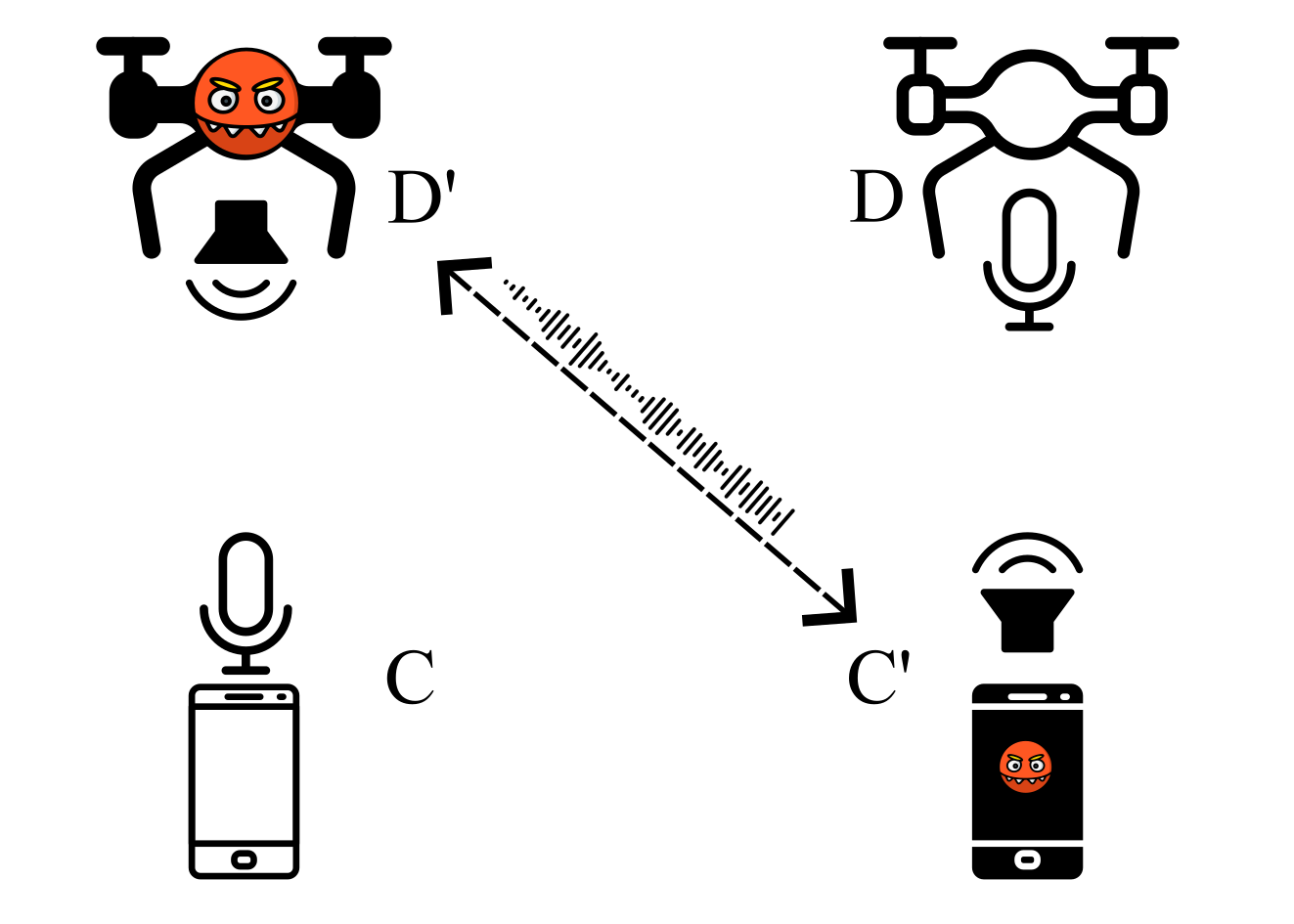}}
\quad\quad\quad\quad
\subfloat[User-side \audiorelay attack]{\includegraphics[scale=0.14]{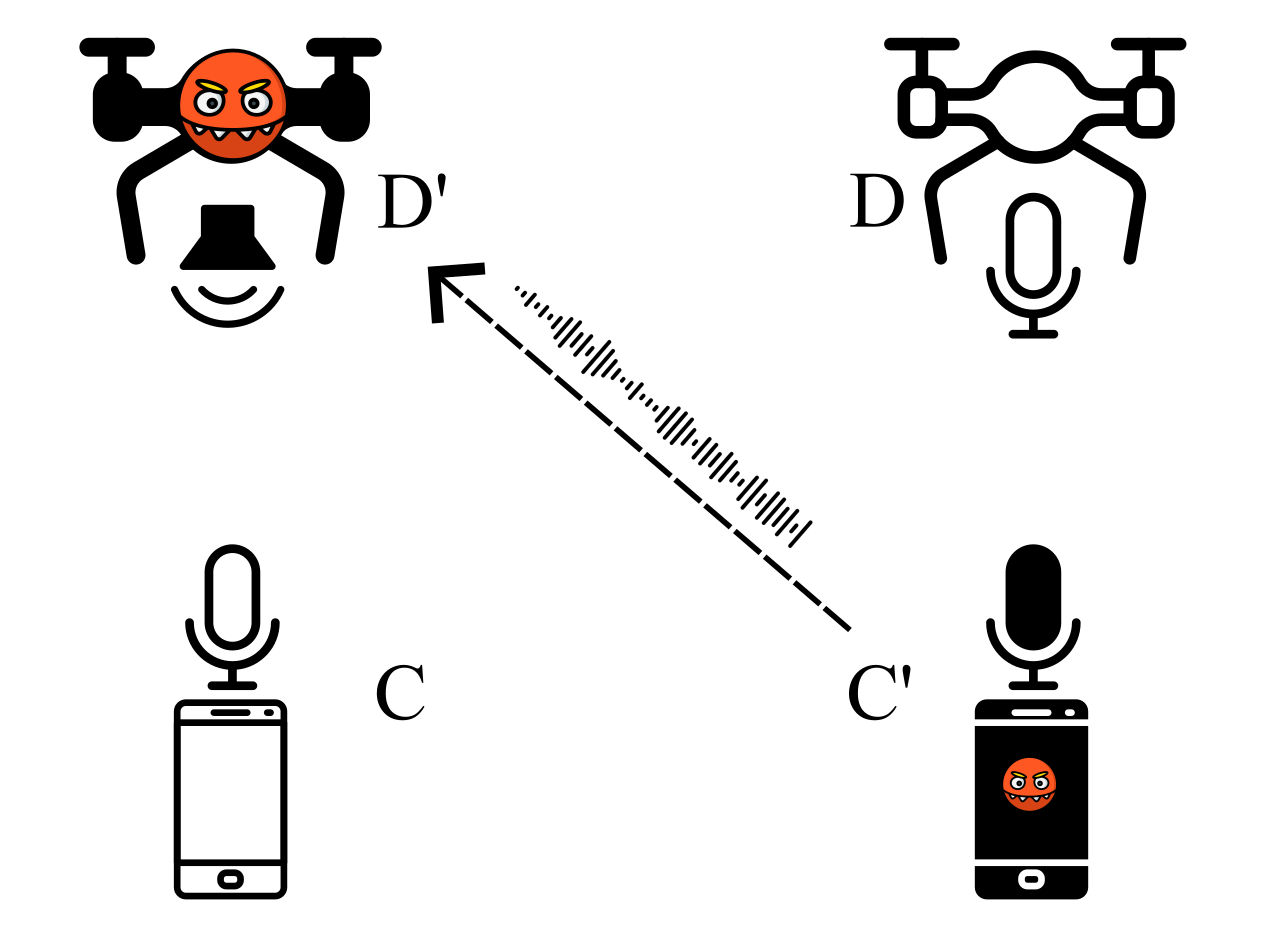}}
\quad\quad\quad\quad
\subfloat[Drone-side \audiorelay attack]{\includegraphics[scale=0.14]{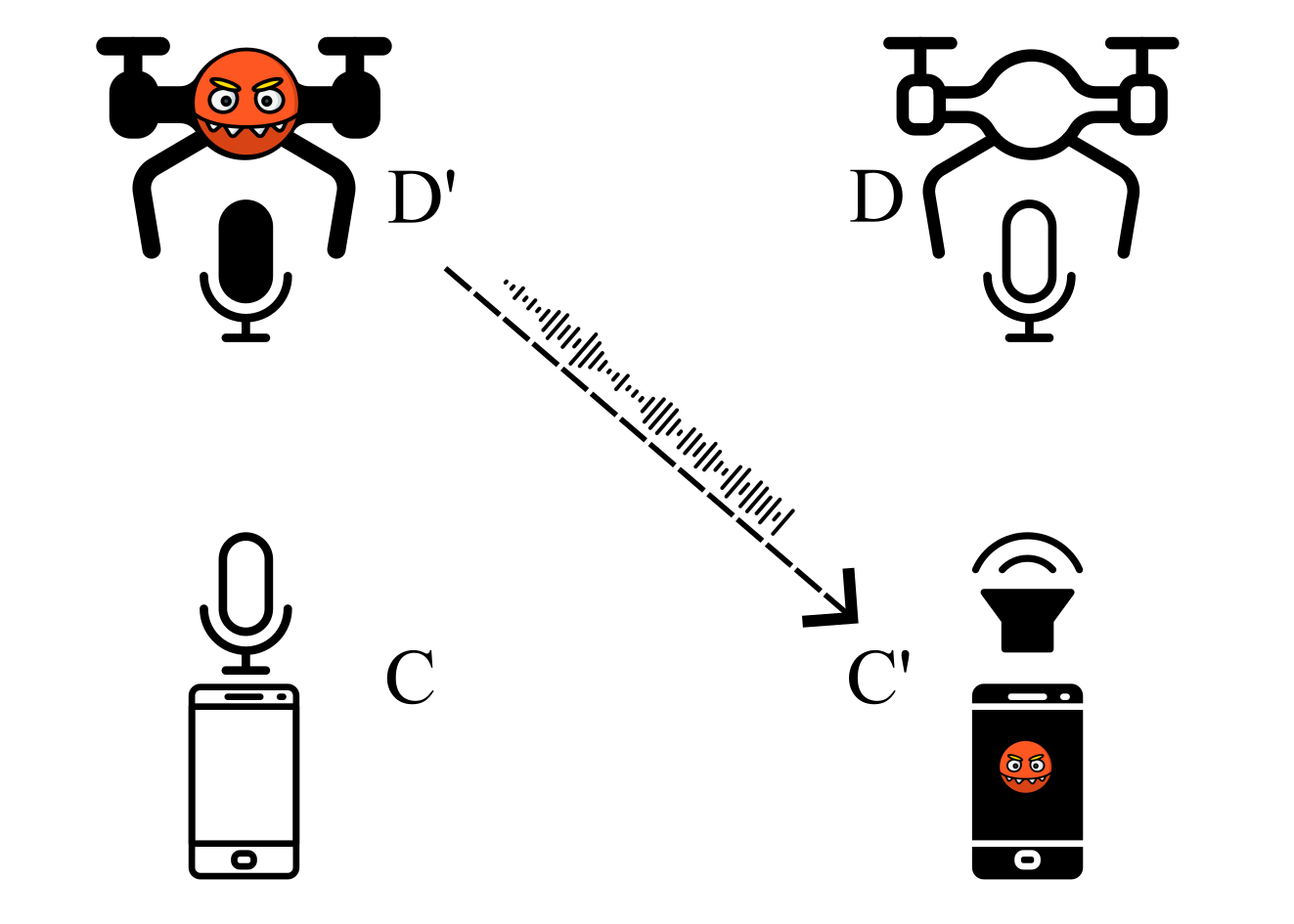}}
\quad\quad
\caption{Various attacks considered in our threat model.
$C$: verifier device; $D$: legitimate drone; $C'$: attacker's device; $D'$: attacker's drone.
}
\label{fig:attacks}
\end{figure*}

\subsection{Threat Model} \label{sub:thread-model}

\Para{(1) Dominant sound attacks.} As shown in Figure~\ref{fig:attacks}(a), an attacker can play identical and synchronized dominant sounds near the service drone and the verifier device, with an intention
that the two devices record similar sounds. (Our evaluation in Section~\ref{Subsec:dominant_attack} demonstrates that when the malicious sound is loud enough, the attack success rate can be over 95\%, as depicted in Figure~\ref{fig:attack_success_rate}(a).)

\para{(2) Audio Relay Attacks.} (i) As shown in Figure~\ref{fig:attacks}(b),
the malicious device records the sounds near $D$ and transmits the recording
over radio to $D'$, which replays it near $C$. We call it a
\emph{verifier-side \audiorelay attack}. 
(ii) Figure~\ref{fig:attacks}(c) shows an attack, where $D'$ records
the sounds near $C$ and transmits the recording to $C'$, which replays it near $D$, called a 
\emph{drone-side \audiorelay attack}.\footnote{We assume the attack has a negligible latency from record-time to replay-time~\cite{zero-latency}. Thus, we do \textbf{not} rely on checking latency to detect attacks.} (Our evaluation in Section~\ref{Subsec:user_side_replay_attack} and Section~\ref{Subsec:drone_side_replay_attack} reveals that the success rate of audio replay attacks can exceed 95\%, as illustrated in Figure~\ref{fig:attack_success_rate}(b) and Figure~\ref{fig:attack_success_rate}(c), respectively.)

\para{(3) Identical Drone Model Attacks.} 
An adaptive attacker can select a malicious drone $D'$,
which is of the same model as the service drone $D$, to fool our system.
Due to manufacturing imperfections, drones make different sounds and noises, even if they are of the same mode. 
Our evaluation in Section~\ref{Subsec:accuracy} finds that 
such attacks has a near-zero attack success rate even when 
a drone hovers. The attack success rate becomes zero when
$D$ randomly and slightly  moves during the authentication time.

In all the attacks, we assume a strong adversary who can use powerful external/directional loudspeakers (i.e., not limited to the built-in loudspeakers of smartphones or drones).

\vspace{3pt}
\noindent\emph{Attacks Out of Scope.} An attacker may launch Denial-of-Service (DoS) attacks, such as {radio jamming~\cite{mpitziopoulos2009survey}}, to disrupt the communication. Handling DoS attacks are beyond the scope
of this work. 

\section{System Overview} \label{Sec:overview}

\subsection{Approach and Assumptions}
\Para{Our Approach.}
Given an order placed by a verifier, who owns the verifier device $C$ (\eg a smartphone), assuming the drone $D$
is dispatched to serve the order; during the authentication procedure,
both $C$ and $D$ record sounds for a short time
(the duration is a parameter studied in Section~\ref{Subsec:parameter_study}). Then, the recordings by $C$ and $D$ are sent to each other  for comparison. (Assuming the drone service company's server
can be trusted,
the computation can be offloaded to the server and the result is
sent to $C$ and $D$.) We are to verify this hypothesis: if $C$ and $D$
are in close proximity, the similarity score of the two recordings
is high; otherwise, low. 

\para{Assumptions.}
We assume a
key-protected
wireless channel between the drone and the verifier, which is assumed in many prior works, 
such as G2Auth~\cite{wu2022g2auth}, Smile2Auth~\cite{sharp2022authentication}, Qualcomm's patent~\cite{qualcomm-code}, distance bounding~\cite{brands1993distance}, and SoundUAV~\cite{sounduav}.
Regarding key establishment, the drone service company's server simply assigns a key unique for each order to the drone and the verifier; or, 
assuming each drone has a digital certificate, the verifier can make use of PKI
for key establishment. It is important to note that despite the utilization of software-level digital certificates to signify the individual identity of each drone, this approach remains susceptible to impersonation attacks. This vulnerability underscores the necessity of drone authentication.

We assume that (i) the verifier device $C$ is installed with the drone service company's app for placing orders and authentication,
and (ii) both $D$ and $C$ have microphones for sound recording, and are not compromised.

\begin{figure}
\af
\includegraphics[scale=0.83]{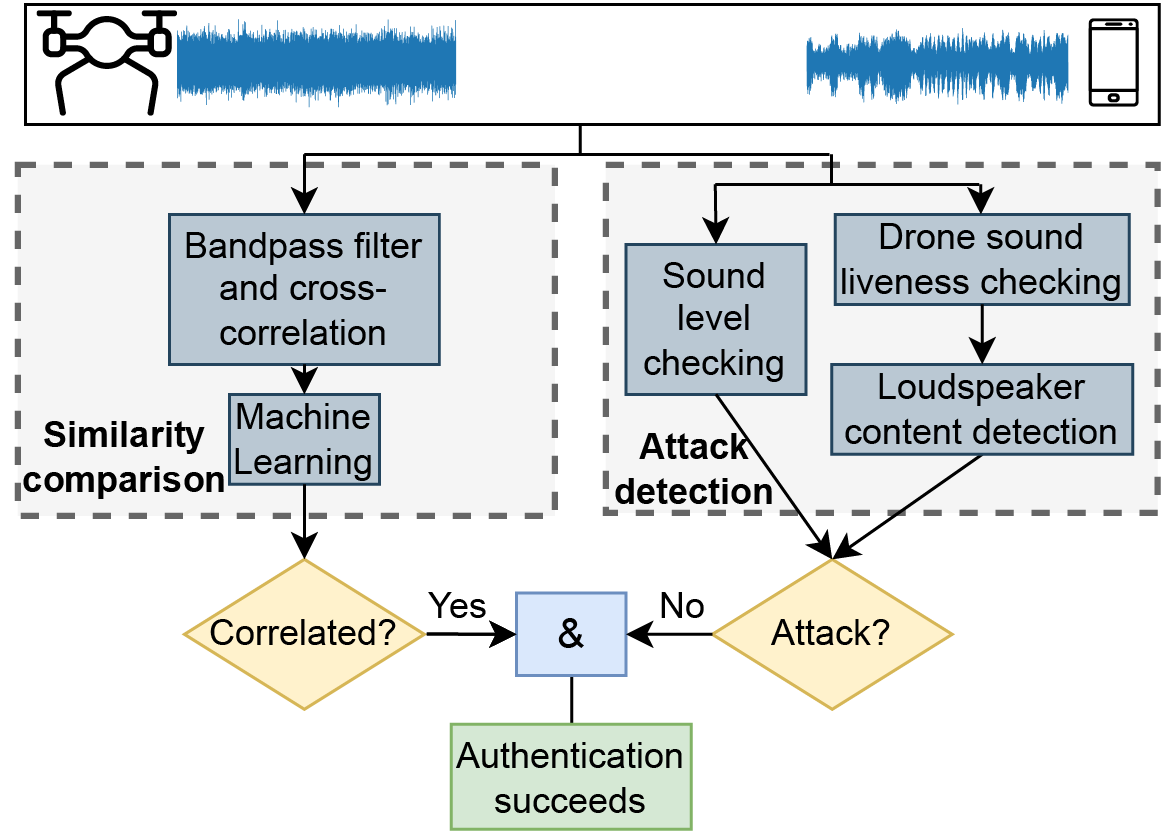}
\caption{Architecture of \shortname.}
\label{fig:architecture}
\end{figure}

\subsection{System Architecture} \label{sec:detailed_design}

To make the discussion concrete, we consider the following \textbf{authentication procedure} between a delivery drone $D$ and a user's smartphone $C$. (The concrete steps may vary when the verifier is a robot or smart doorbell.)

\begin{enumerate} 
    \item After  the drone $D$ arrives, it hovers and establishes a key-protected communication channel with the user's smartphone $C$.      
    Next, $C$ vibrates to let the user $U$ know the delivery drone's arrival.
    
    \item $U$ then walks to $D$ and unlocks $C$ to confirm that she is near $D$ (note the drone hovering in front
    of her may be a malicious one, and
    $D$ may hover near an attacker due to navigation inaccuracy~\cite{gps_accuracy} or GPS signal spoofing~\cite{kerns2014unmanned, shepard2012evaluation,shen2020drift,zeng2018all}; 
    we thus need authentication). 
    
    \item  
    $D$ and $C$ record audios 
    for a short duration $T$, which is studied as a parameter in our evaluation.
    
    \item $D$ and $C$ exchange the recordings to calculate a similarity score. If the authentication succeeds, the drone service proceeds; otherwise, it goes back to Step 3 until the maximum number of attempts is reached.

\end{enumerate}

As shown in Figure~\ref{fig:architecture}, given the two recordings from $D$ and $C$,  \shortname 
performs similarity comparison  and attack detection. 
The \emph{\textbf{similarity comparison}} module (Section~\ref{subsec:authentication}) extracts features
from both the time and frequency domains, which are fed into
a machine learning model to make a decision whether the
two recordings are similar.
The \emph{\textbf{attack detection}} module (Section~\ref{sec:attack-detection}) not only detects the various
attacks in our Threat Model but also considers environmental
sounds that may cause inaccuracies. 
In short, the authentication succeeds only if the two recordings are similar \textbf{and}
no attacks are detected. 

\para{Non-human scenarios.} It is worth highlighting that our authentication approach does not rely on
humans. For example, in the drone delivery scenario, a hovering drone and a dock equipped with
a microphone can apply \shortname for mutual authentication. 
When a drone hovers near a destination dock,
it sends a notification to the dock to start audio recording. 
Similarly, a smart doorbell 
containing a microphone can also conduct mutual authentication with a drone. 
Therefore, \shortname has prominent
\textbf{advantages} over prior state of the art~\cite{wu2022g2auth,sharp2022authentication} that relies on human interactions to conduct
drone authentication: (1) \emph{Usability}: Very often, a user is not home or does not want to wait for the drone service \eg drone delivery. Or, during a very cold or hot day, a user
does not want to go outside to interact with a drone and would rather use her smart
doorbell to finish authentication. 
(2) \emph{Privacy}: Some users do not want to show themselves in front of a drone, which usually carries a camera. (3) \emph{Generalizability}: \shortname extends its applicability to other scenarios, including the authentication process between a drone and a robot/door bell/warehouse.

\subsection{Multiple Drones and Verifiers}
\para{Multiple drones.}
When multiple drones hover near $C$, even if the authentication result is positive, it is difficult for $C$ to decide which drone is the right one. 
Note prior state of the art~\cite{wu2022g2auth,sharp2022authentication} also needs to tackle this issue.
The verifier can update the destination location slightly for a single-drone space.
If another drone follows closely to the new location,  it reveals an attack explicitly.
Another simplest solution is that the service drone waits until it is the only drone nearby to
conduct the authentication. An attacker may control a drone to stay for DoS
attacks; however, since the attacker cannot make a profit but explicitly reveals the malicious drone, 
such attacks are unlikely to be attractive to attackers.

\para{Multiple verifiers.} 
A positive authentication result can 
only tell the legitimate verifier is in close proximity to the legitimate drone. 
When there are multiple verifiers nearby, e.g., in a popular place, the service drone $D$ cannot decide which verifier is the right one. Sound waves in the range 16kHz to 20kHz  are near-ultrasound insensitive to humans~\cite{ka2016near}, and loudspeakers can play sounds in this range~\cite{near-ultrasound}. Thus, $D$ can request
$C$ announce a sequence of specified numbers in this frequency
range, and $D$ conducts sound source localization~\cite{valin2003robust}.
A device nearby that clones the behavior reveals itself as a malicious one
but cannot make a profit.

\section{Studying Drone Noises}
\label{subsec:preliminary_study}

\begin{figure}
\center
\subfloat[No payload]{\includegraphics[scale=0.265]{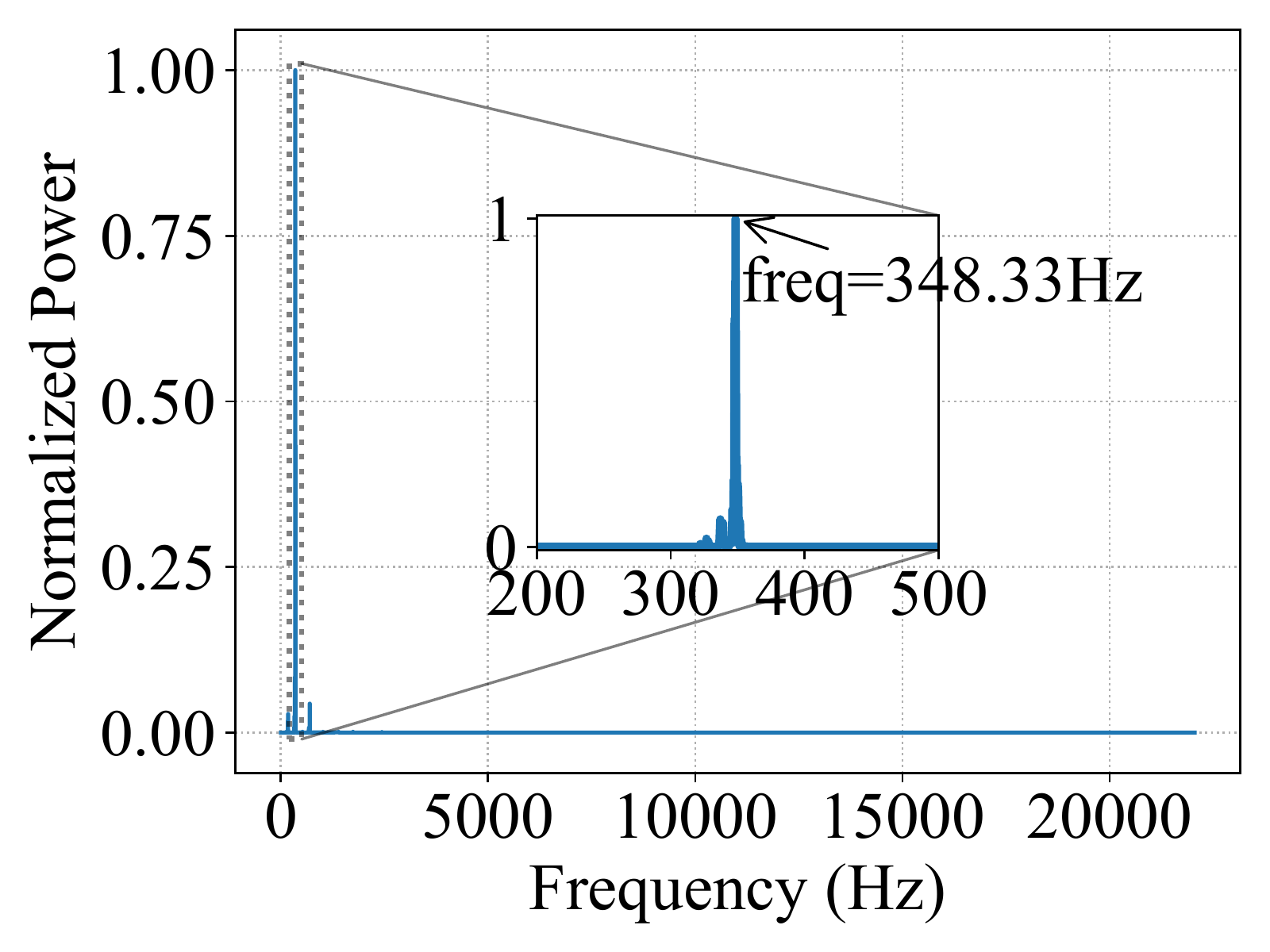}}
\subfloat[Full payload]{\includegraphics[scale=0.265]{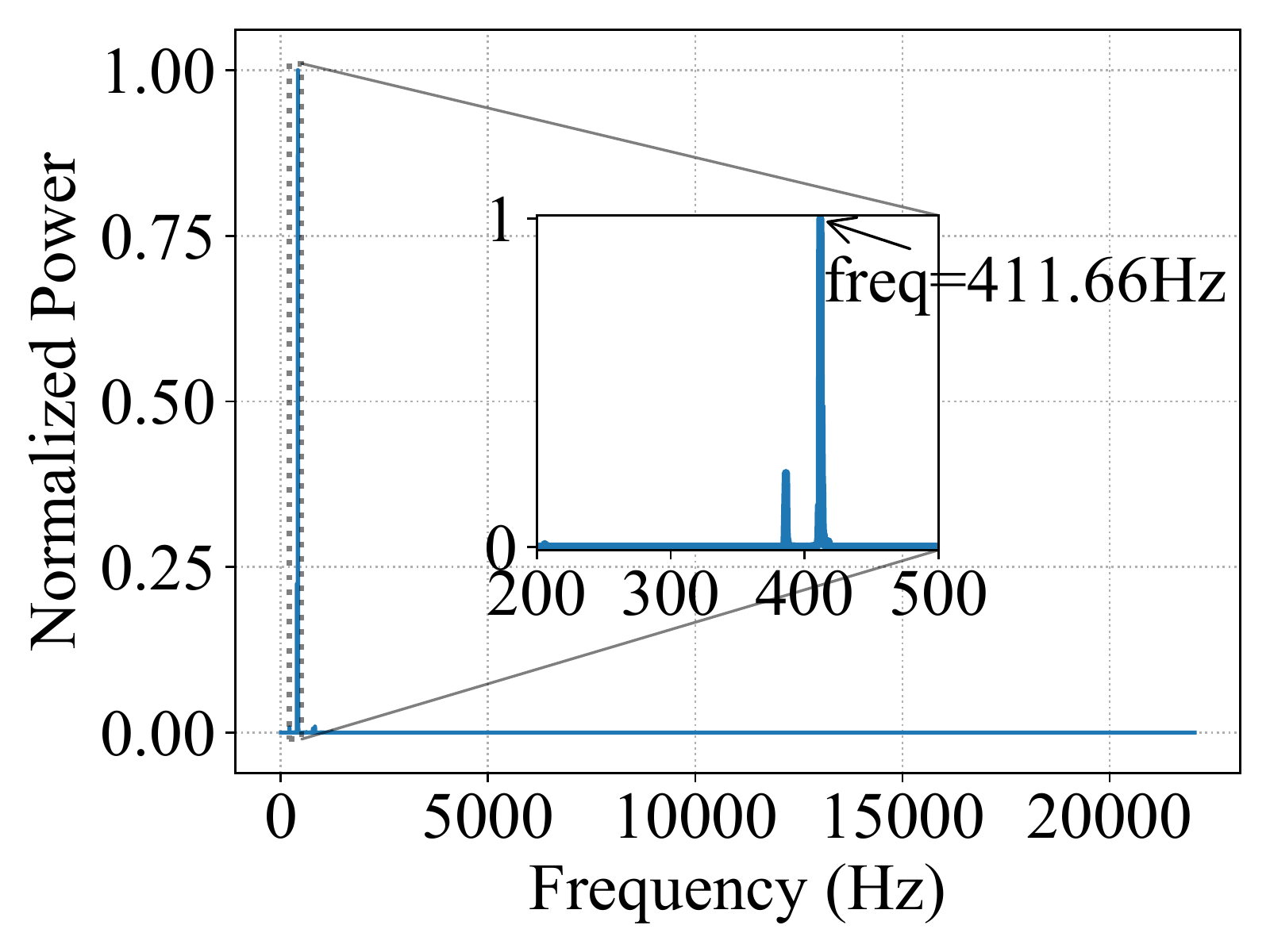}}
\caption{Power spectral density of sounds made by the same drone with different payloads.}
\label{fig:spectrum_payload}
\end{figure}

When a drone hovers, the spinning blades push air down and meanwhile the air pushes up on the rotors, lifting the drone. It is worth noting that when a drone hovers, it needs to continuously
adjust its actuators slightly to keep itself balanced~\cite{azinheira2008hover}. 
The sounds made by the propellers and the motors of a drone are related to their rotational rates, which vary due to many factors, such as the air flowing around the drone, the payload carried by the drone, the current acceleration, and the posture change of the drone~\cite{tinney2018multirotor}.

For example, when a drone hovers with heavier payloads, to remain steady, the drone needs a greater lift force. Thus, the blades spin faster, which is reflected by the \emph{\textbf{essential
frequency}}---the frequency that has the \emph{greatest} amplitude.
We conducted an experiment to study the sounds of a Mavic Mini drone. Figure~\ref{fig:spectrum_payload} shows how the essential frequency varies with different payloads. When the drone hovers with no payloads, the essential frequency of the sounds is 348.33Hz; in comparison, when the drone carries full payloads, it increases to 411.66Hz. 

\begin{figure}
\centering
\includegraphics[scale=0.35]{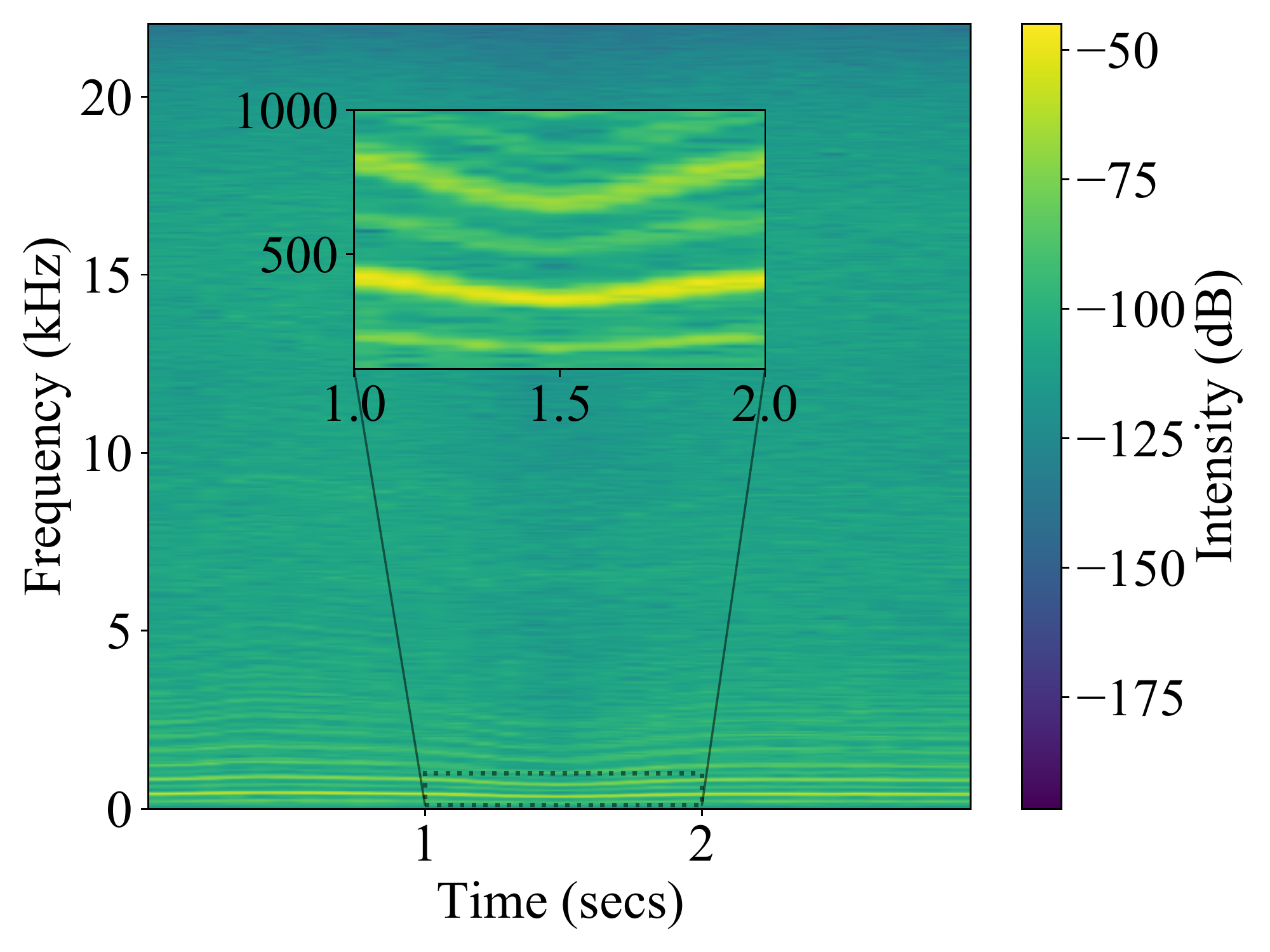}
\aaf\aaf
\caption{Spectrogram of sounds made by a hovering drone. The repetitive stripes show harmonics.}
\label{fig:stripes}
\end{figure}

Figure~\ref{fig:stripes} shows how the essential frequency changes in another experiment. From 1s to 1.5s, the drone reduces the rotation speed of its blades, which makes the drone 
drop down slightly and the essential frequency decrease; from 1.5s to 2s, the drone raises the rotation speed, making the drone fly up and the essential frequency increase. (Regardless of the payloads and operations, however, we find \emph{the essential frequency lies in the low-frequency range}, which is consistent with prior studies~\cite{shi2018anti, bannis2020bleep}.)

Given the multiple factors that complicate drone noises, 
we seek a \emph{fingerprint-free} authentication method.
Moreover, as the air flow is complex in nature and hard to predict, plus multiple other factors, such as  
unique manufacturing imperfections in the motors, it is unlikely to reproduce
the noises of one drone using another, even when they are of the same model, which is verified (Section~\ref{Sec:Evaluation}). 

\section{Sound Similarity Comparison}
\label{subsec:authentication}

\subsection{Background: Cross-correlation}
\label{subsubsec:cross_correlation}

Cross-correlation serves as a widely-utilized standard method for measuring similarity, as demonstrated in previous applications~\cite{sound-proof}. It tracks the movement of one or more sets of time series relative to one another, which can be used to determine how well they match up with each other.
Assuming two independent time series of length $n$ are denoted as $X$ and $Y$, cross correlation $c_{XY}(l)$ with a lag $l$ can be calculated as:
\aaf
\begin{equation}
c_{XY}(l)=\sum_{i=0}^{n-1} X(i)Y(i-l)
\end{equation}
where $Y(i)=0$ if $i<0$ or $i>n-1$.
To accommodate for different amplitudes of the two signals, the cross correlation can be normalized as: $c'_{XY}(l)=\frac{c_{XY}(l)}{\sqrt{c_{XX}(l)c_{YY}(l)}}$. The normalized cross correlation 
has the range of [-1, 1], where 1 indicates the two time series have the same shape; -1 indicates the two time series have the same shape but with opposite signs; and 0 indicates they are uncorrelated. 

The computation overhead 
can be optimized by using the Fourier transformation: 
$c_{XY}(l)=F^{-1}(F(X)^*\cdot F(Y))$, where $F()$ denotes the Fourier transformation, $F^{-1}()$ denotes the inverse Fourier transformation, and the asterisk denotes the complex conjugate. 

\subsection{Exploiting Drone Noises: Essential Frequency-Centered Feature Selection}
\label{subsubsec:one_third_bands}

\begin{figure}
\center
\subfloat[Waveform]{\includegraphics[scale=0.207]{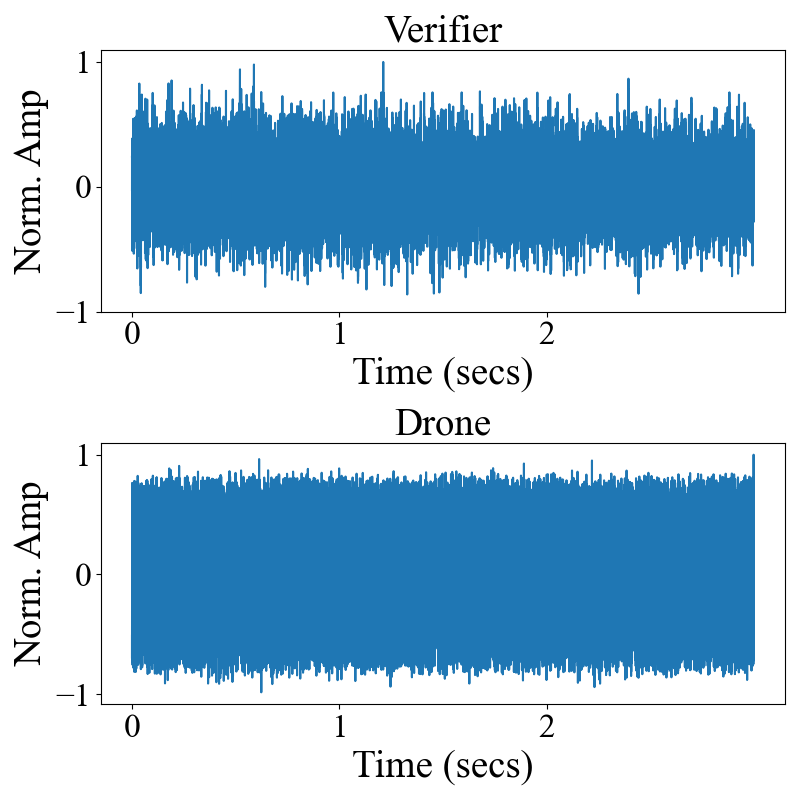}}
\subfloat[Spectrum]{\includegraphics[scale=0.207]{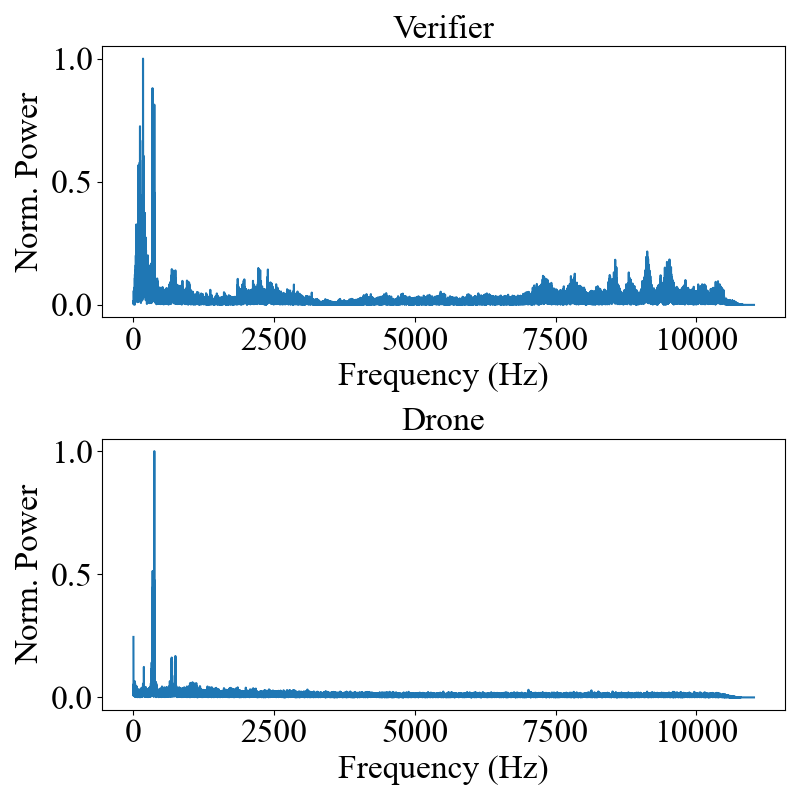}}
\caption{Comparison of audios recorded by the verifier and the drone.}
\label{fig:audio_comparison}
\aaf
\end{figure}

\Para{Raw features.}
The raw audio recordings from the verifier side and the drone side exhibit significant differences due to varying distances from the drone. As depicted in Figure~\ref{fig:audio_comparison}, the waveforms and spectra of the recordings from both sides exhibit noticeable disparities in shape, posing challenges in assessing their similarity directly from the raw data.
To measure similarity in a fine-grained scale, \emph{bandpass filters}~\cite{standard2004s1} can be
used to divide the sounds into \emph{multiple} frequency bands and the maximum cross-correlation value in each band can be used as a similarity metric. 
Specifically, the audible range of frequencies are divided in one-third octave bands, which
split the first 10 octave bands in three and the last octave band in two, for a total of 32 bands. One-third octave bands are widely used in acoustics and their frequency ranges have been standardized~\cite{standard2004s1}.  
After calculating the maximum cross-correlation value in each of the bands, a 
vector of 32 dimensions is derived, each dimension denoting the cross-correlation value in one band.

\para{Insight and investigation.}
Instead of regarding  all the 32 bands as equal, 
our insight is that the bands are \emph{not} equal, and that the essential frequencies of a drone usually lie in a narrow low-frequency range (see Section~\ref{subsec:preliminary_study}). 

\begin{figure}
\centering
\includegraphics[scale=0.35]{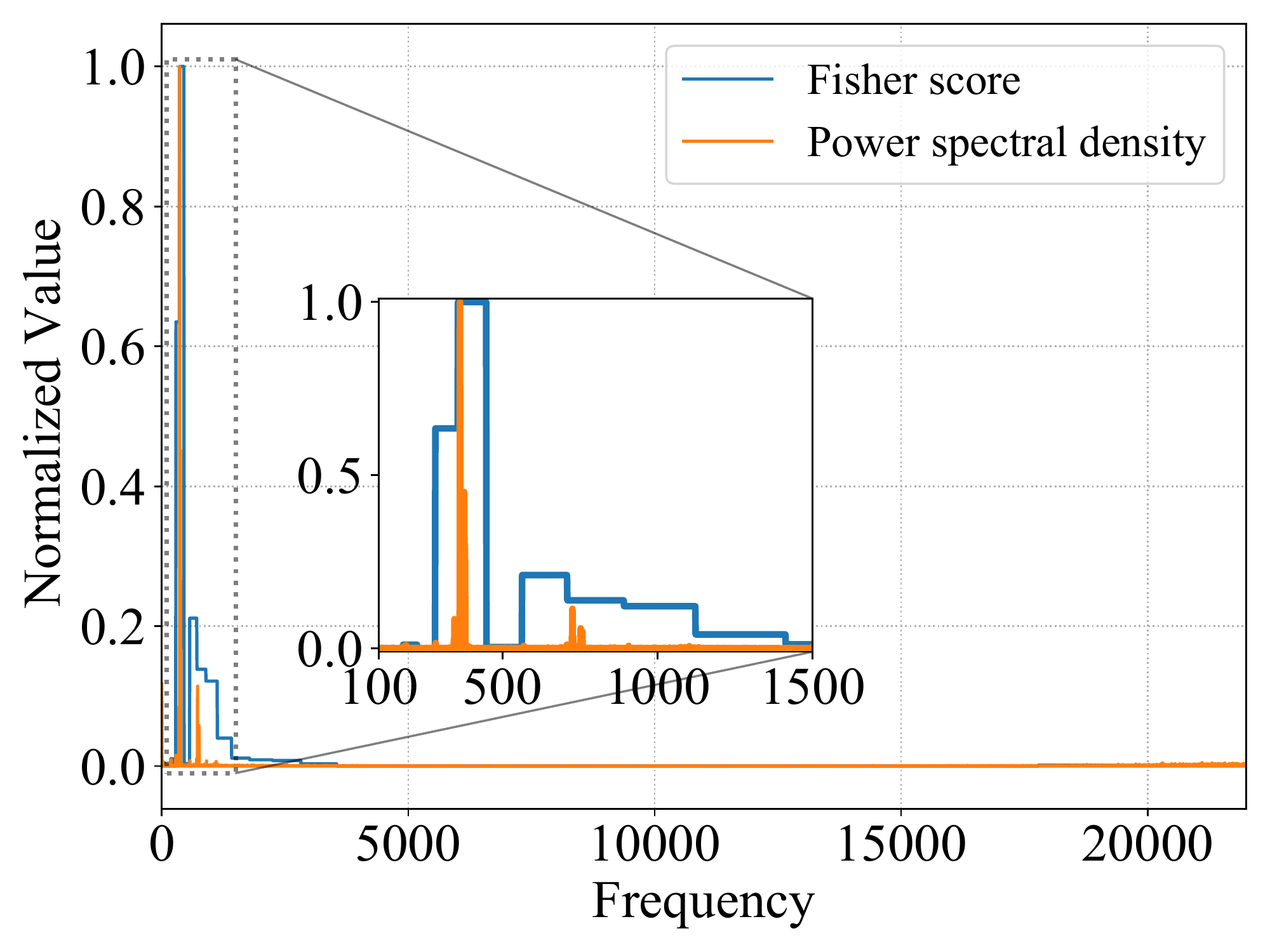}
\aaf\aaf
\caption{Fisher scores and power spectral density.}
\label{fig:fisehr_score}
\aaf\af
\end{figure}
 
We then apply a Fisher 
scoring algorithm~\cite{li2018feature}, which is
used for feature selection, to study the importance of the scores from different bands. It ranks the features prior to a learning task.  
We compute the normalized Fisher 
scores of all the features with the data points collected in our dataset
(detailed in Section~\ref{subsubsec:dataset1}), and compare the Fisher 
scores with the power spectral density of the sound. Figure~\ref{fig:fisehr_score} shows the results of normalized Fisher 
scores and power spectral density over all the frequency bands. The maximum Fisher 
score is from the frequency band [355Hz, 447Hz], which covers the essential frequencies of the drone. 
Figure~\ref{fig:fisehr_score} also reveals that \emph{features in frequencies with higher power spectral density tend to have higher scores}, which interprets and confirms the
validity of Fisher scores.
 
\para{Idea.} 
This inspires us to propose that \emph{the fine-grained bandpass filters should take a closer look at the frequencies with high power spectral density}. For example, Figure~\ref{fig:fisehr_score} shows that the power of sound made by a hovering drone is concentrated around its essential frequency. Therefore, instead of using the standard one-third octave bands~\cite{standard2004s1}, we propose to divide the frequency band   that covers the essential frequencies of a drone (the frequency width is denoted as $N$ Hz) into $M$ equal parts, each occupying $N/M$ Hz. The value of $N$
should take into consideration the range of essential frequencies when a drone carries different weights of payloads. (Our parameter study finds $M=5$ works well across different drones.)

\shortname discards all other bands and  only uses information from the $M$ features, called \emph{\textbf{essential frequency-centered feature selection}}, to train a binary classifier, 
which has the following benefits:
(1) it excludes the interference of a large variety of environmental sounds;
and 
(2) it significantly reduces the computational overhead. 

\section{Defeating Attacks \& Tolerating Environmental Sounds} \label{sec:attack-detection}
\subsection{Main Idea} \label{subsec:attack-detection-idea} 
\Para{Detecting dominant sound attacks and drone-side \audiorelay attacks.}
Due to the very loud drone noises and the very small distance between a drone's sound sources
and its microphone,
the  microphone of a drone mainly records drone noises. 
(Also note that when there are attacks, $C$ is close to $D'$ 
and $C$ mainly records the drone noises of $D'$.) Attackers who launch 
dominant sound attacks (Figure~\ref{fig:attacks}(a)) or
drone-side \audiorelay attacks (Figure~\ref{fig:attacks}(c)) 
need to impose \textbf{very} loud malicious sounds on $D$ to distort the recording by $D$ (in order to obtain a high similarity score), which
motivates us to check the sound level on the  side of $D$ to detect the two kinds of attacks (Section~\ref{subsubsec:sound_level_checking}).

\para{Detecting verifier-side \audiorelay attacks.}
A verifier-side \audiorelay attack (Figure~\ref{fig:attacks}(b))  
records the noises of $D$ and replays them towards $C$. As this attack does not
play any sounds towards $D$, the detection idea above does not work.
To make the attack stealthy, the attacker may use a directional loudspeaker. We propose to check the
liveness of the sounds recorded by $C$; that is, whether the sounds are due to a live drone or a
loudspeaker (Section~\ref{subsubsec:speaker_detection}). 

\para{Reducing false rejections.}
(1) Naive sound level checking may cause false rejections, as environmental sounds, such as traffic noises, can also influence the sound level. 
(2) What complicates the liveness detection is that there may be loudspeakers playing music or news on the verifier side. 
Not only does our work defeat the various attacks that are ignored by prior work~\cite{sounduav,daaf}, we also
devise methods to tolerate various environmental sounds to reduce false rejections.

\subsection{Sound Level Checking}
\label{subsubsec:sound_level_checking}
Our observation is that to have a significant impact on the sounds recorded by $D$, an attacker has to play malicious sounds loudly. We thus study the feasibility of detecting such attacks based on sound level checking. 

A sound level can be denoted as the intensity ($I$) of the sound wave, which is the amount of energy at a given area per unit of time. It is measured as the intensity level ($IL$) in decibel scale (dB), which can be converted from the sound intensity with the unit of $W/m^{2}$ following the formula $IL=10log(I/10^{-12})$~\cite{fahy2017sound}. The intensity of combined sounds is the sum of the individual intensities due to the independent sound sources~\cite{lighthill1954sound}. As a result, the sound intensity $I'_{D}$ measured by the microphone on the drone's side  can be denoted as $I'_{D}=I_{D}+I_{E}$, where $I_{D}$ is the sound intensity of the drone noises and $I_{E}$ the environmental noises. 

To significantly distort the noises recorded by the drone-side microphone, an attacker ($A$) has to make sounds of intensities above $I_{Amin}$. Assuming the typical environmental ($E$) sound level is not greater than  $I_{Emax}$ and $I_{Emax}<I_{Amin}$, a threshold of $I'_{D}$, which is between $I'_{Dlow}=I_{D}+I_{Emax}$ and $I'_{Dhigh}=I_{D}+I_{Amin}$, can be used to detect attackers while allowing benign verifiers to be authenticated under typical environmental sounds, such as traffic. 
Typical environmental sounds are below 95 dB.\footnote{It is damaging to hearing after long exposure to sounds above 95 dB and it can cause hearing loss after exposure to sounds above 100 dB~\cite{cdc_hearing}.} Thus, we consider $I_{Emax}=95$ dB.

On the other hand, a drone spins its motors and propellers intensively when hovering, which makes loud noises. 
The sound level measured by the microphone on the drone side is thus very high. 
For example, the sound level measured by a  
microphone attached to 
the DJI Mavic Mini drone
is $99.3 \pm 1.8$ dB.  
According to our evaluation,  an attacker has to play sounds above $I_{Amin}$=100 dB to get a high similarity score. Based on the formulas above, $I'_{Dlow}=100.672$ dB and $I'_{Dhigh}=102.674$ dB. The threshold  $I'_D$ on the drone side can be chosen between $I'_{Dlow}$ and $I'_{Dhigh}$, 
such that \shortname impedes attacks by checking whether the recorded sound level is above the threshold, and its tolerate environmental sounds below 95 dB; i.e., it does not cause false rejections. To measure sound levels during authentication,  drones can be equipped with inexpensive sound level meters. For example, a small decibel detection module, which costs \$12.25, can measure sounds between 40 dB and 130 dB with a resolution of 0.1 dB~\cite{decible_measure_module}.

Our evaluation (Section~\ref{Sec:security_analysis}) examines different drones, malicious sounds, and environmental sounds.

\subsection{Drone Sound Liveness Checking and Loudspeaker Content Detection}
\label{subsubsec:speaker_detection}

We have the two observations: (1) to launch a successful verifier-side audio relay attack, a loudspeaker
needs to replay a drone's sounds on the verifier side; 
and (2) a legitimate verifier may be playing music
and news using a loudspeaker. 

Therefore, if no loudspeakers are detected (that is, sound liveness checking passes), we are
ensured that there are no verifier-side audio relay attacks. Otherwise, 
we do \emph{not} 
immediately reject the authentication: if the loudspeaker plays sounds, such as music and
news, it should not be reported as attacks.
(However, if the verifier happens to be playing drone sounds from a loudspeaker, a false rejection will occur. Such corner cases are not common 
and \shortname can generate an alert to ask the verifier to pause them.)
Because of hardware imperfections, sounds produced by a loudspeaker exhibit distinct characteristics compared to original sounds. These characteristics have been used to differentiate between live-human voices and replayed ones by prior studies~\cite{ahmed2020void,lavrentyeva2017audio,blue2018hello}. Expanding on this notion, we propose to utilize these inherent loudspeaker features not only for sound liveness checking but also for identifying whether the sounds emitted by loudspeakers are drone noises. Notably, while loudspeakers need to produce loud sounds to imitate drone noises, their distortion tends to increase at high amplitudes~\cite{klippel2006tutorial}. As a result, drone sounds played by loudspeakers are distinguishable from authentic drone noises.
We assess the following features to study the two questions, including (i) \emph{decay patterns} in spectral power, (ii) \emph{peak patterns} in spectral power, and (iii) \emph{linear prediction cepstrum coefficients} (LPCC). Since these features stem from inherent hardware imperfections of loudspeakers, there currently exist no known attacks capable of circumventing sound liveness checking.

\begin{figure}
\graphicspath{ {./Figures/} }
\centering
\includegraphics[scale=0.26]{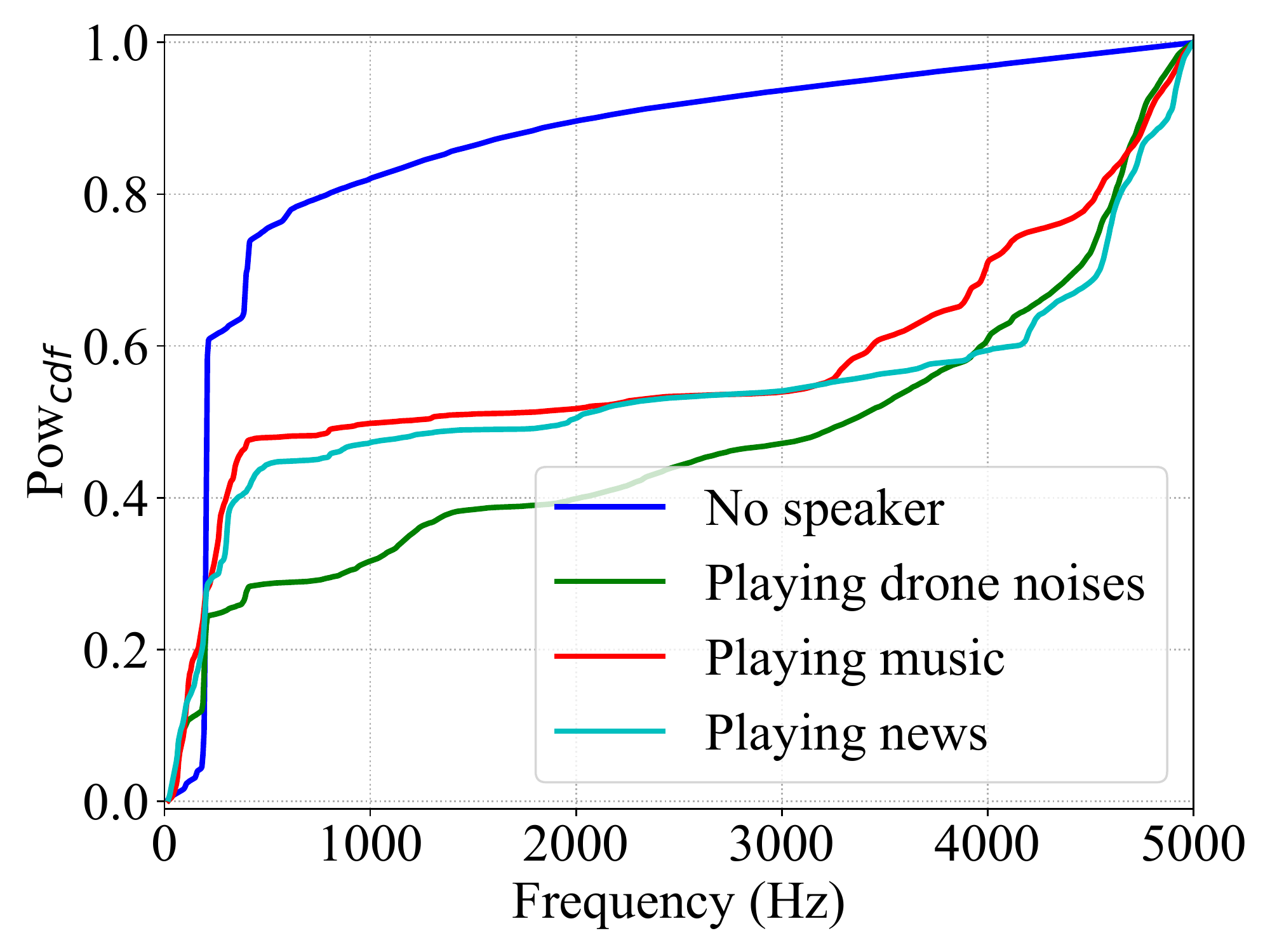}
\caption{Cumulative distribution of spectral power density.}
\label{fig:speaker_cdf}
\end{figure}

To compare the \textbf{decay patterns} of audios recorded in different scenarios, we first record the live sounds made by a drone. Then, we fly a drone of the same model to act the role of a malicious drone $D'$ and meanwhile use a loudspeaker to play the recorded drone noises, random music, or news.  Figure~\ref{fig:speaker_cdf} shows an example cumulative distribution of spectral power density for the sounds recorded  by
the verifier device $C$ in different scenarios. For the live-drone sounds (the ``No speaker'' line), the cumulative power density increases sharply in the frequency below 500Hz, and about 75\% of the total power lies in it. However, in the loudspeaker cases, no matter which sound is played by the loudspeaker, less than 50\% of the total power lies in the frequency below 500Hz; in addition, the cumulative spectral power density increases slowly within the frequency ranging from 500Hz to 4.5KHz, and about 75\% of the total power lies in the frequencies below 4.5KHz. Thus, this feature can well distinguish live sounds from loudspeaker sounds.

\begin{figure}
\graphicspath{ {./Figures/} }
\centering
\includegraphics[scale=0.33]{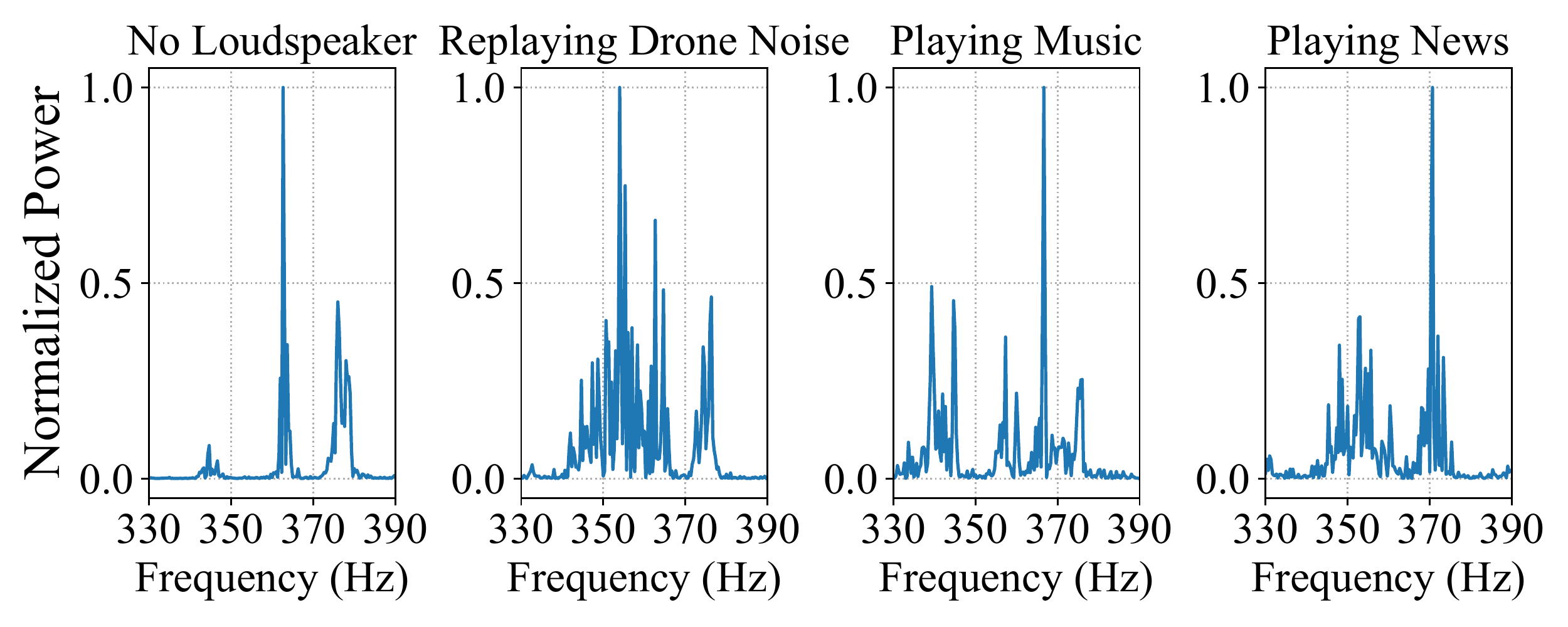}
\caption{Signal power spectrum.}
\label{fig:speaker_power}
\end{figure}

We also examine the \textbf{peak patterns} in spectral power of the sounds generated in the scenarios 
above. 
Figure~\ref{fig:speaker_power} shows the power spectrum of the sounds collected above around the essential frequency. It can be seen that the power of live drone sounds is more concentrated around its essential frequency with fewer peaks, while the sounds recorded in
the scenario of replaying the recorded sounds have the most  fluctuations and different sounds show
different fluctuations. 

The sounds produced by a real drone hovering in the air has a very narrow dominant frequency range, while the sounds played by loudspeakers have much wider frequency ranges. 
The decay and peak patterns mainly look at low-frequency ranges.
To better distinguish loudspeakers playing music and news from those playing drone's noises, we perform an 
examination of wider frequency ranges. \textbf{LPCC} uses the energy values of linear filter banks, which equally emphasize the contribution of all frequency components of an audio. 
Thus, LPCC is chosen as a complementary feature to help cover frequencies in wider ranges. 

In short, we extract the following classification features: low frequencies power features, higher power frequencies features, signal power linearity degree features, and LPCC features. With these features, we train two 
machine learning based binary classifiers: one is to distinguish whether the sounds are due to a loudspeaker, and the other whether the loudspeaker is playing drone noises.

\section{Evaluation}
\label{Sec:Evaluation}

We received an IRB approval. An extensive evaluation is conducted to
study these questions.
\textbf{Q1}: How accurate is \shortname (Section~\ref{Subsec:accuracy})?
\textbf{Q2}: How do different parameters affect \shortname (Section~\ref{Subsec:parameter_study})?
\textbf{Q3}: How resilient is \shortname to attacks and how robust is it to environmental sounds (Section~\ref{Sec:security_analysis})?
\textbf{Q4}: How quick is the authentication (Section~\ref{Subsec:efficiency})?

\subsection{Experimental Setup and Data Collection}
\label{Subsec:data_collection}
\subsubsection{Devices and Metrics}
\label{subsubsec:experiment_setup}

\begin{figure}
\graphicspath{ {./Figures/} }
\centering
\includegraphics[width=\linewidth]{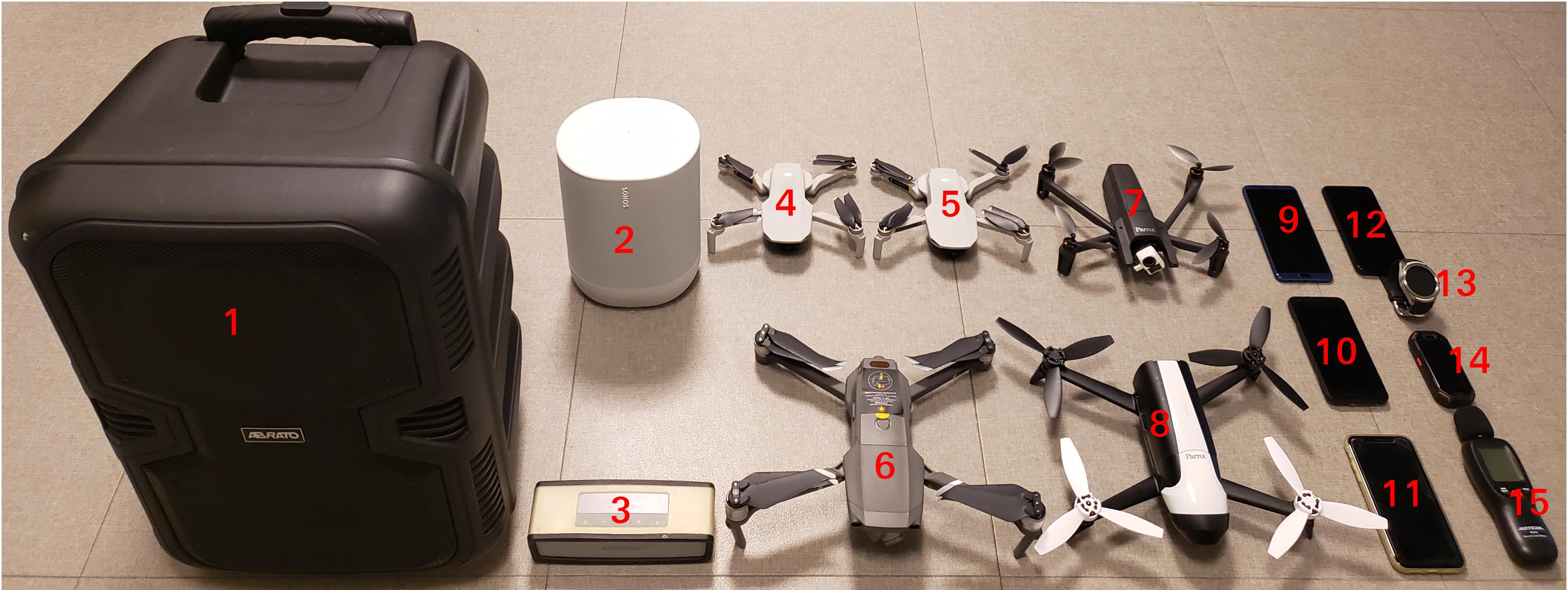}
\caption{Devices used in our experiments: three speakers: an Abrato loudspeaker, labeled as 1, a SONOS loudspeaker as 2, and a Bose loudspeaker as 3; five drones: two DJI Mavic Mini drones labeled as 4 and 5, a DJI Mavic 2 Zoom drone as 6, a Parrot ANAFI Thermal drone as 7, and a Parrot Bebop 2 drone as 8; five smartphones: an Honor V10 labeled as 9, a Nexus 5X as 10, an iPhone 11 as 11, a Pixel 4 as 12, and a Unihertz Atom as 14; an LG smartwatch W200 labeled as 13; a Meterk sound level meter labeled as 15.}
\label{fig:devices}
\end{figure}

\Para{Devices.}
We use a variety of devices to collect data. 
As shown in Figure~\ref{fig:devices}, five (5) drones are used: two DJI Mavic Mini drones, a DJI Mavic 2 Zoom drone, a Parrot ANAFI Thermal drone, and a Parrot Bebop 2 drone. A small Android smartphone, Unihertz Atom, is closely attached to the drones to record drone noises. Four (4) smartphones, an Honor V10, a Nexus 5X, an iPhone 11, a Google Pixel 4, and a smartwatch LG W200 are used on
the verifier side for sound recording. 
Three loudspeakers (Abrato, SONOS, and Bose) positioned at a distance of 4 meters from the drones and verifiers are used to play sounds and launch attacks. We use a MacBook Pro with a 2.6 GHz dual-core Intel Core i5 processor 
as the server for data processing and decision making.

\para{Metrics.}
\emph{False Acceptance Rate}~(FAR) and \emph{False Rejection Rate}~(FRR) are two metrics used to evaluate the accuracy of \shortname. FAR denotes the percentage of instances where attacks are incorrectly authenticated and a lower FAR indicates higher security. FRR shows the percentage of legitimate authentication instances are rejected and a lower FRR shows better usability. We also report \emph{Equal Error Rate} (EER) when FRR is equal with FAR. A lower EER indicates a higher accuracy of a system. We also use a \emph{Receiver Operating Characteristics}~(ROC) curve to show the accuracy of our system across all possible thresholds and report \emph{Area Under the Curve} (AUC) of the ROC curve.

\subsubsection{Dataset I for Accuracy Evaluation}
\label{subsubsec:dataset1}
To evaluate the accuracy of our system, we build \emph{Dataset~I}. 
\textbf{All the 5 drones are used} for this purpose. For each drone, we hover it at a 
height of 4 meters and place the 
smartphone 5 meters vertically apart from the drone. 
In each authentication process, the smartphone and the drone record audios through their microphones for 3 seconds. Note the parameters, such as the distance and 
the recording duration, are studied in Section~\ref{Subsec:parameter_study}.

\para{Positive pairs.} When a drone hovers in front of a smartphone, the smartphone and the drone record audios simultaneously. The two audios form a positive data pair $(c, d)$, where $c$ is the audio recorded by the smartphone and $d$ by the drone. For each drone, we collect 1000 positive data pairs, each labeled as $l=1$.

\para{Negative pairs.} 
Assuming we have two positive data pairs $(c1, d1)$ and $(c2, d2)$ generated using two drones $D1$ and $D2$, respectively, we generate two negative pairs: $(c1, d2)$ and $(c2, d1)$. For each drone, we generate 250 negative data pairs with each of the other drones, so that each drone has 1000 ($= 250\times 4$) negative data pairs, each labeled as $l=0$.

\subsubsection{Dataset II for Attack Resilience Evaluation}
\label{seubsubsec:dataset2}
Two DJI Mavic Mini drones, \ie $D$ and $D'$, are selected to build Dataset II as they are of the same model (hence, presumably generating the most similar drone sounds and meaning the highest possible attack success rate) and generate the least loud noises among the drones in our experiments (hence, needing attack sounds at the lowest sound level). $D$ acts as the service drone and $D'$ acts as the malicious one. 
The audios recorded by $D$, $D'$, $C$, and $C'$ 
are denoted as $d$, $d'$, $c$, and $c'$, respectively.

\para{Dataset II-A.} 
To assess resilience to dominant sound attacks (Figure~\ref{fig:attacks}(a)), we  build \emph{Dataset II-A}. We first record the sounds made by a drone.
As shown in Figure~\ref{fig:attacks}(a), when $D'$ hovers near $C$ (and $D$
near $C'$), we play 
the recorded drone sounds loudly via an external loudspeaker, and have the smartphones and the drones record audios simultaneously. 
(Note that 
the loudspeaker is placed 
to keep the same distance, 4 meters, to drone and smartphone, so that the malicious sound level measured on both sides are \emph{equal}.)
We change the malicious sound level, measured on the phone/drone side,
from 95 dB to 115 dB in steps of 5 dB and record 1000 data pairs at each sound level. Data pairs, $(c, d)$ and $(c', d')$, are used to build this dataset.

\para{Dataset II-B.}
To evaluate resilience to verifier-side \audiorelay attacks, we build 
\emph{Dataset II-B}. As shown in Figure~\ref{fig:attacks}(b), the audio, denoted as $d$, produced and recorded by $D$ is replayed near $C$ via a loudspeaker. 
Data pairs, each in the form of $(c, d)$, constitute the dataset.  
We place a loudspeaker 4 meters away from the smartphone and change the volume of the loudspeaker, which is measured on the phone side, from 55 dB to 75 dB
in steps of 5 dB and record 1000 data pairs at each sound level. 

\para{Dataset II-C.}
To evaluate resilience to drone-side \audiorelay attacks, we build \emph{Dataset II-C}. As shown in Figure~\ref{fig:attacks}(c), we record the noises produced by $D'$ and replay it where $D$ hovers via a loudspeaker placed 4 meters away. The audio $d$ indicates the sound recorded by drone $D$ under drone-side \audiorelay attack. 
Thus, data pairs, each in the form of $(c, d)$, are used to build this dataset.
We change the volume of the loudspeaker, which is measured on the \emph{drone side}, from 95 dB to 115 dB in steps of 5 dB and record 1000 data pairs at each sound level.

\af
\subsection{Authentication Accuracy}
\label{Subsec:accuracy}

\begin{figure}
\graphicspath{ {./Figures/} }
\centering
\includegraphics[scale=0.32]{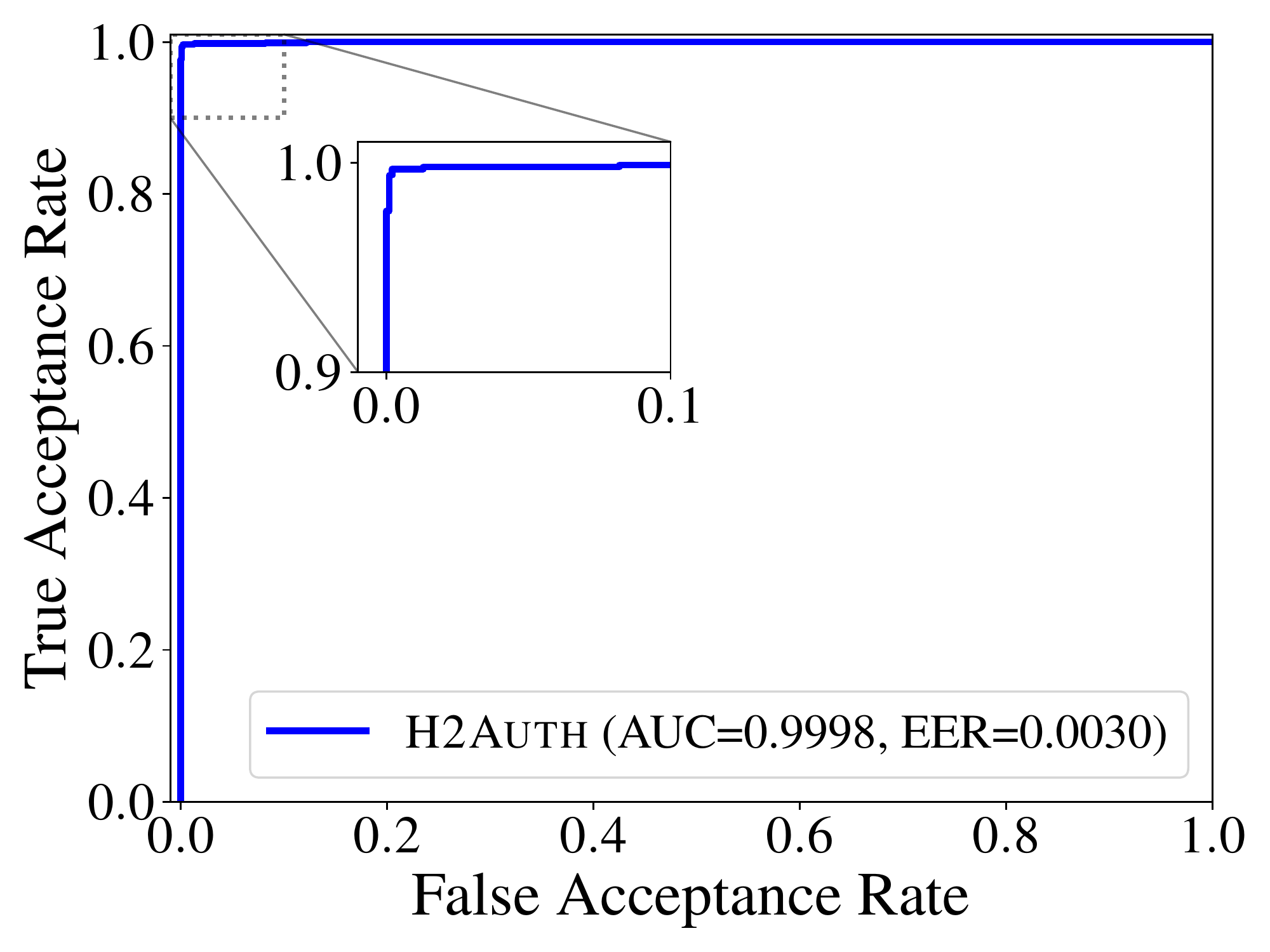}
\aaf\af
\caption{ROC curve, AUC and EER.}
\label{fig:roc_and_far_frr}
\aaf\af
\end{figure}

\Para{Results.} We use Dataset I and the 10-fold cross validation to train and test our SVM model (the classifier is studied
as a parameter in Section~\ref{Subsec:parameter_study}). Figure~\ref{fig:roc_and_far_frr} shows the ROC curve,  
EER=0.0030 and AUC=0.9998. 
The low EER indicates that \shortname has a very high \textbf{accuracy} ($=1-$EER) 0.9970. 
Both \shortname and the prior work Acoustic Fingerprint~\cite{daaf} do not need to land to conduct
authentication, while Acoustic Fingerprint relies on fingerprints and its accuracy is 96.2\%.

We further evaluate how secure \shortname is
under \textbf{Identical Drone Model Attacks}.
We compare the sounds of $D'$ recorded by a verifier-side smartphone
with the sounds of $D$ recorded by $D$ itself. 
We construct 1000 such negative data pairs using two 
Mavic Mini drones of the same model. 
The results show that only 0.6\% of the negative samples are classified as positive, indicating the high accuracy of our system in distinguishing drones of the same model. Moreover, when $D$ randomly
and slightly moves around, the attack success rate becomes zero.

\begin{figure}
\center
\subfloat[With gusty winds]{\includegraphics[scale=0.21]{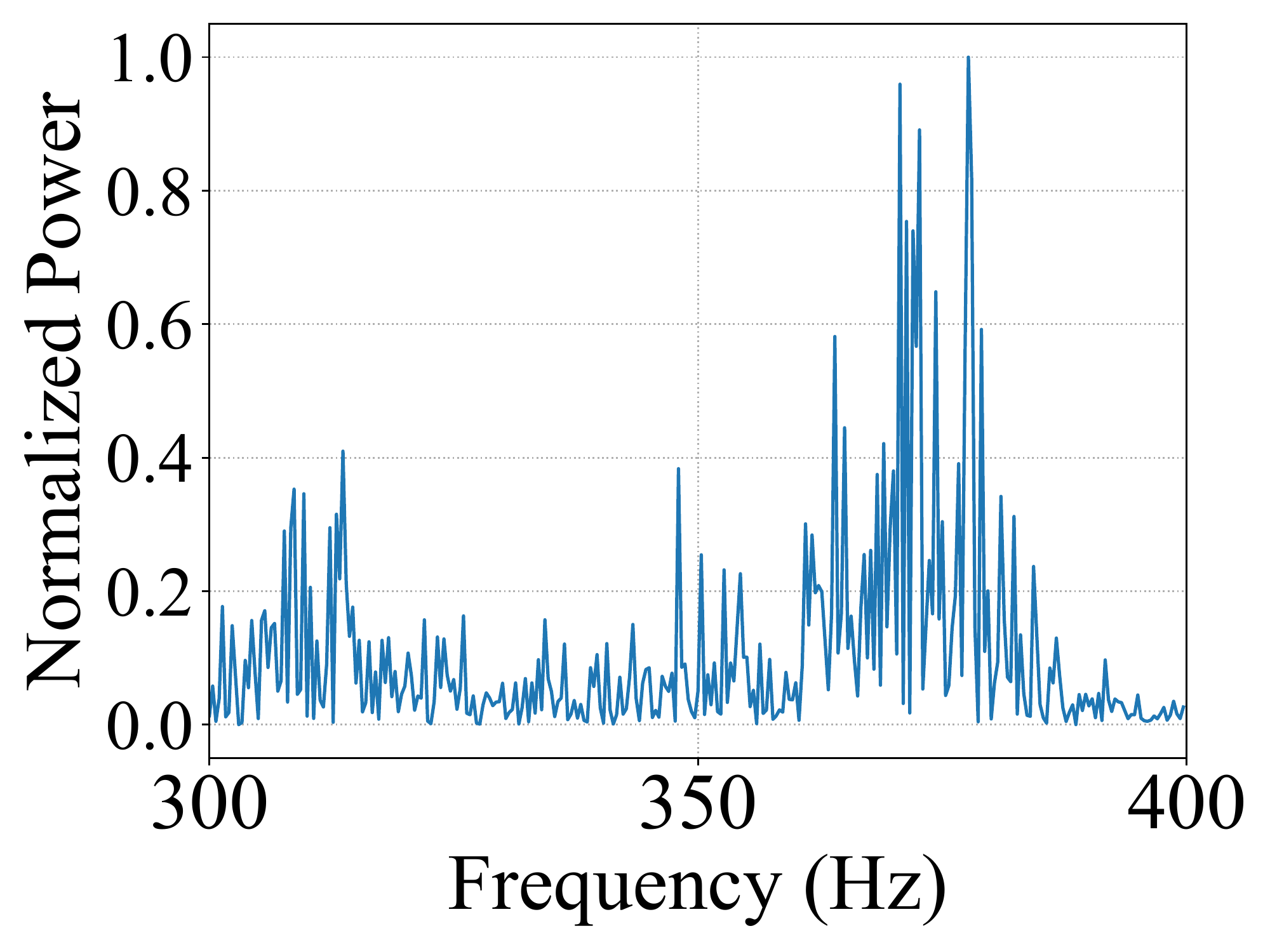}}
\subfloat[Without gusty winds]{\includegraphics[scale=0.21]{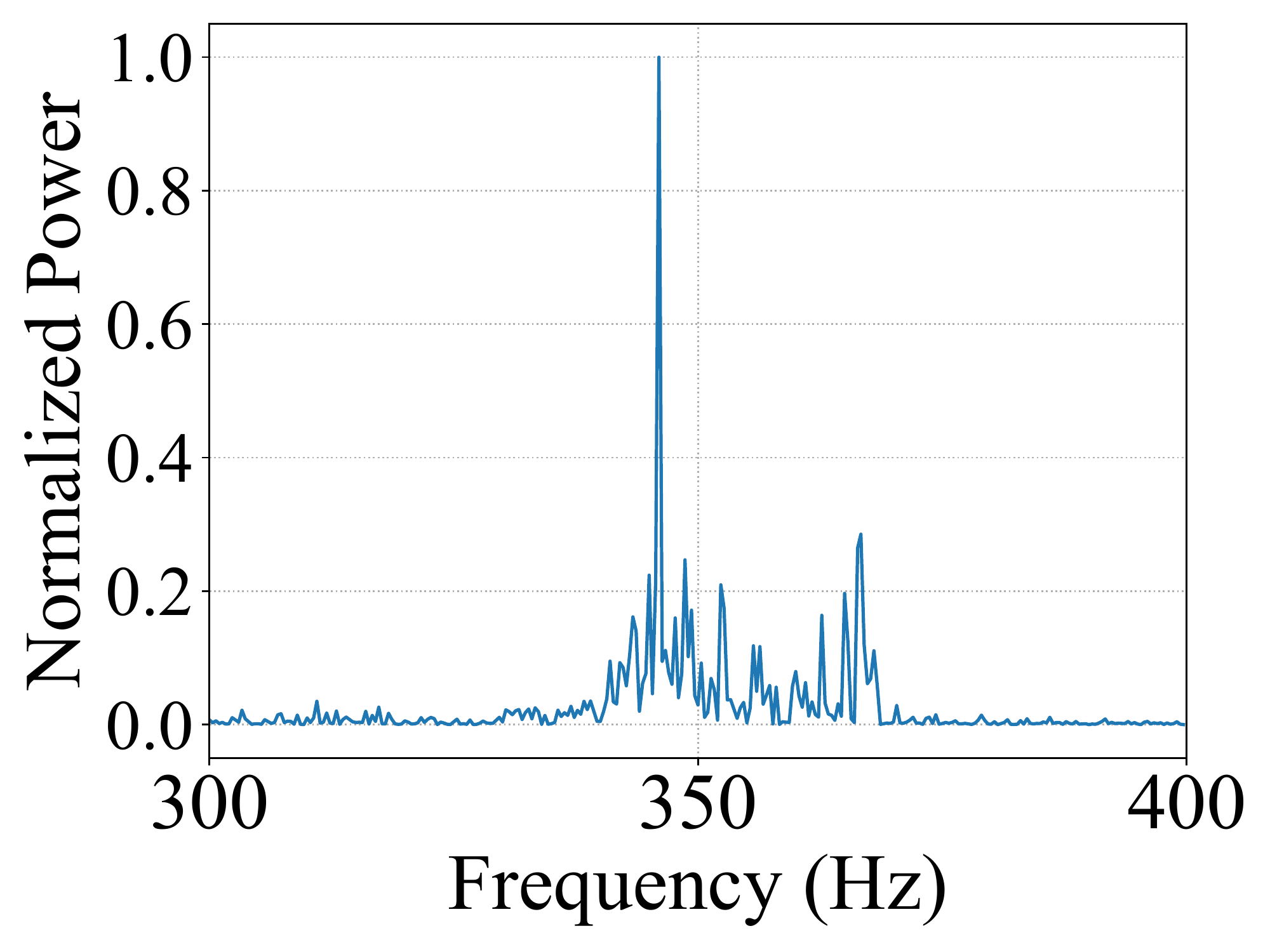}}
\aaf
\caption{Power spectrum of sounds recorded by the drone in different wind conditions.}
\label{fig:spectrum_wind}
\aaf\af
\end{figure}

\para{Failure analysis.} The rare failure cases are caused by occasional gusty winds, as noises increase on microphones of the hovering drone with gusty winds~\cite{walker2010review}. Figure~\ref{fig:spectrum_wind} shows the power spectrum of the sounds recorded on the drone side on different wind conditions, where Figure~\ref{fig:spectrum_wind}~(a) is a failure case and Figure~\ref{fig:spectrum_wind}~(b) a regular case. The gusty winds rush across the microphone attached to the drone, which induced noises directly. 
To mitigate the impact caused by gusty winds, we find it effective to add a windscreen to cover the microphone~\cite{hosier1979microphone}. We thus suggest service drones carry windscreens on the microphones. 

\subsection{Parameter Study}
\label{Subsec:parameter_study}

\begin{figure*} 
\aaf\aaf
\subfloat[EER \emph{vs.} training dataset size]{\includegraphics[scale=0.28]{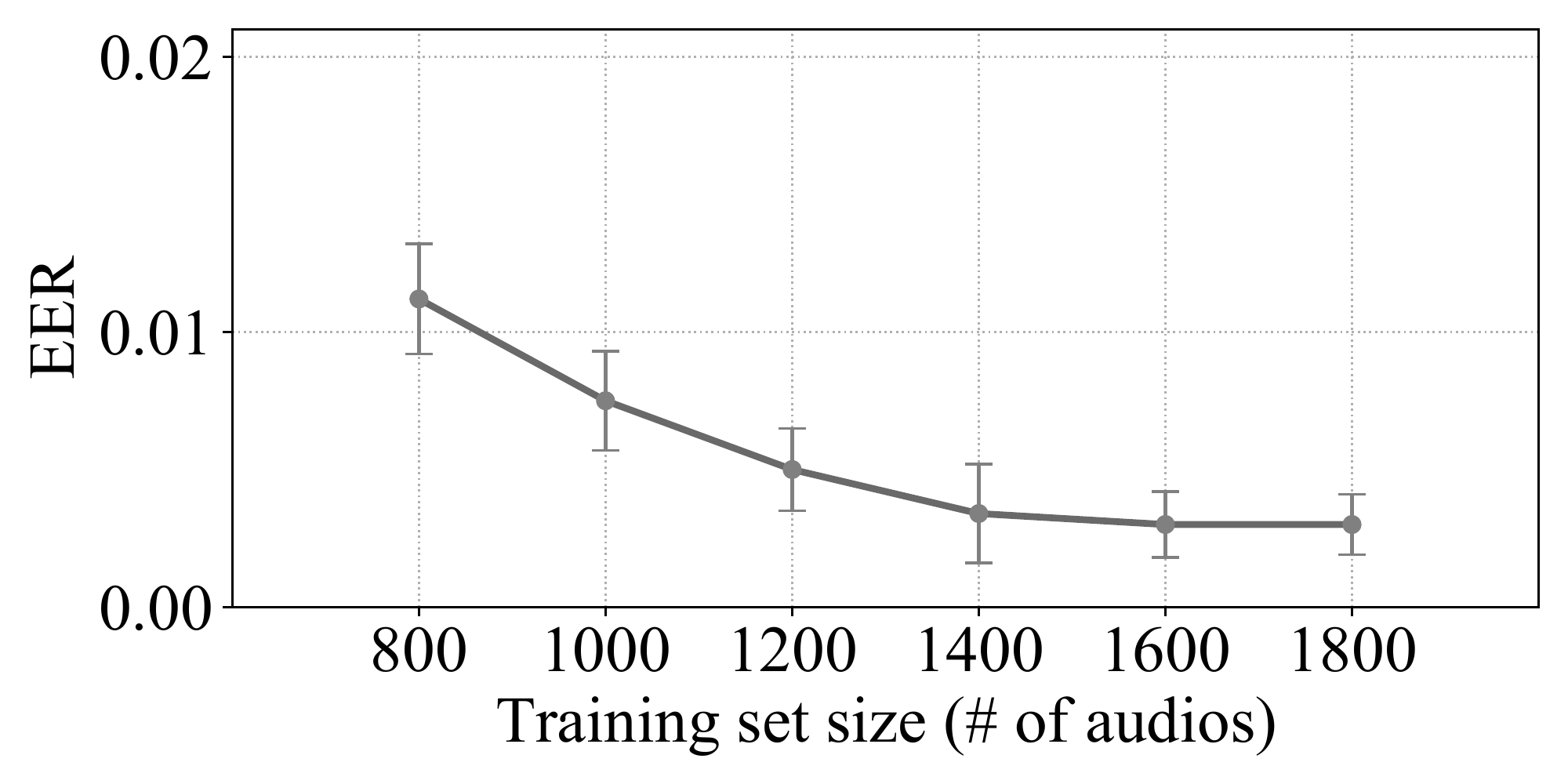}} ~ 
\subfloat[EER \emph{vs.} distance]{\includegraphics[scale=0.28]{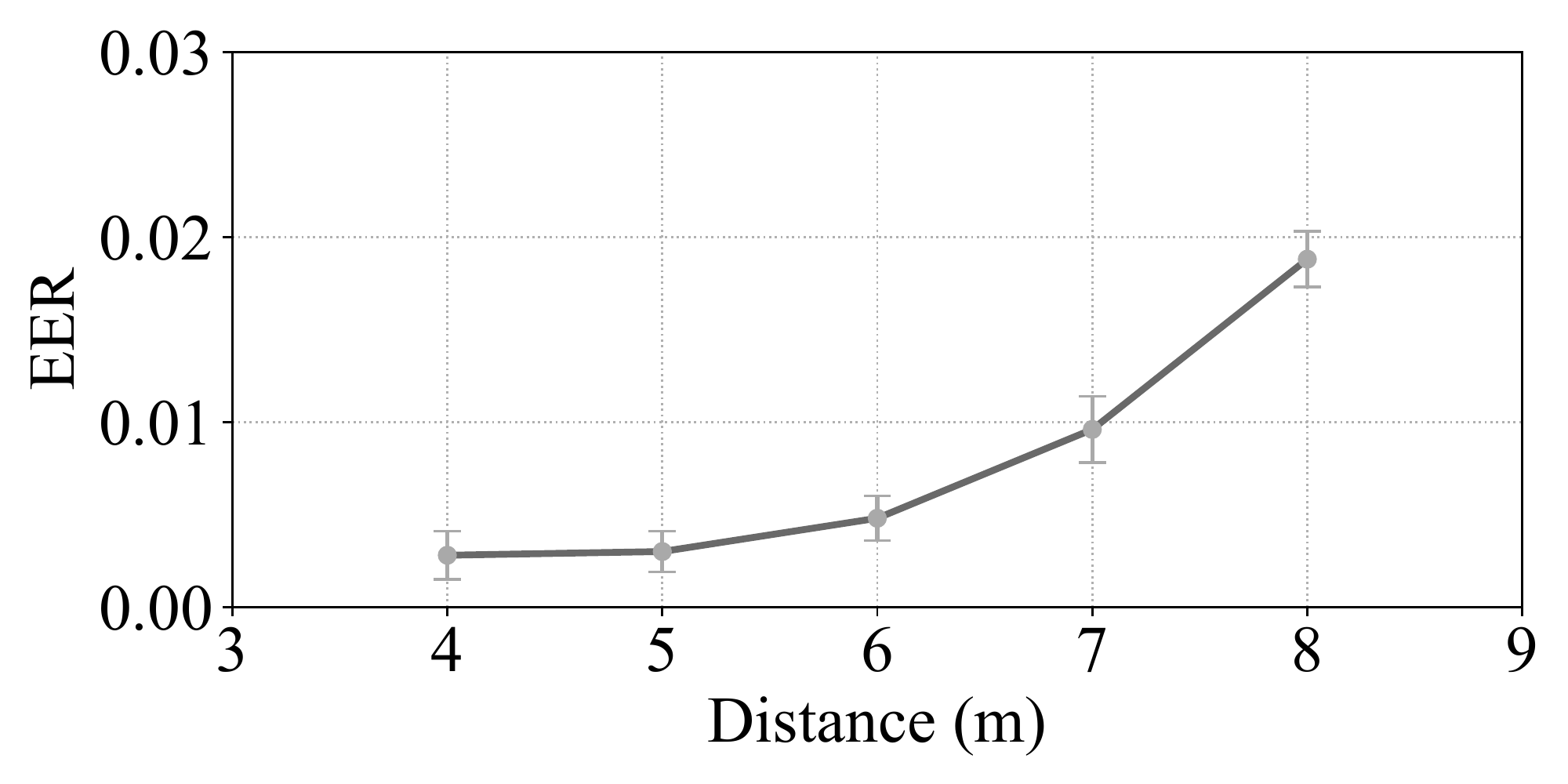}}~
\subfloat[EER \emph{vs.} environmental sounds]{\includegraphics[scale=0.28]{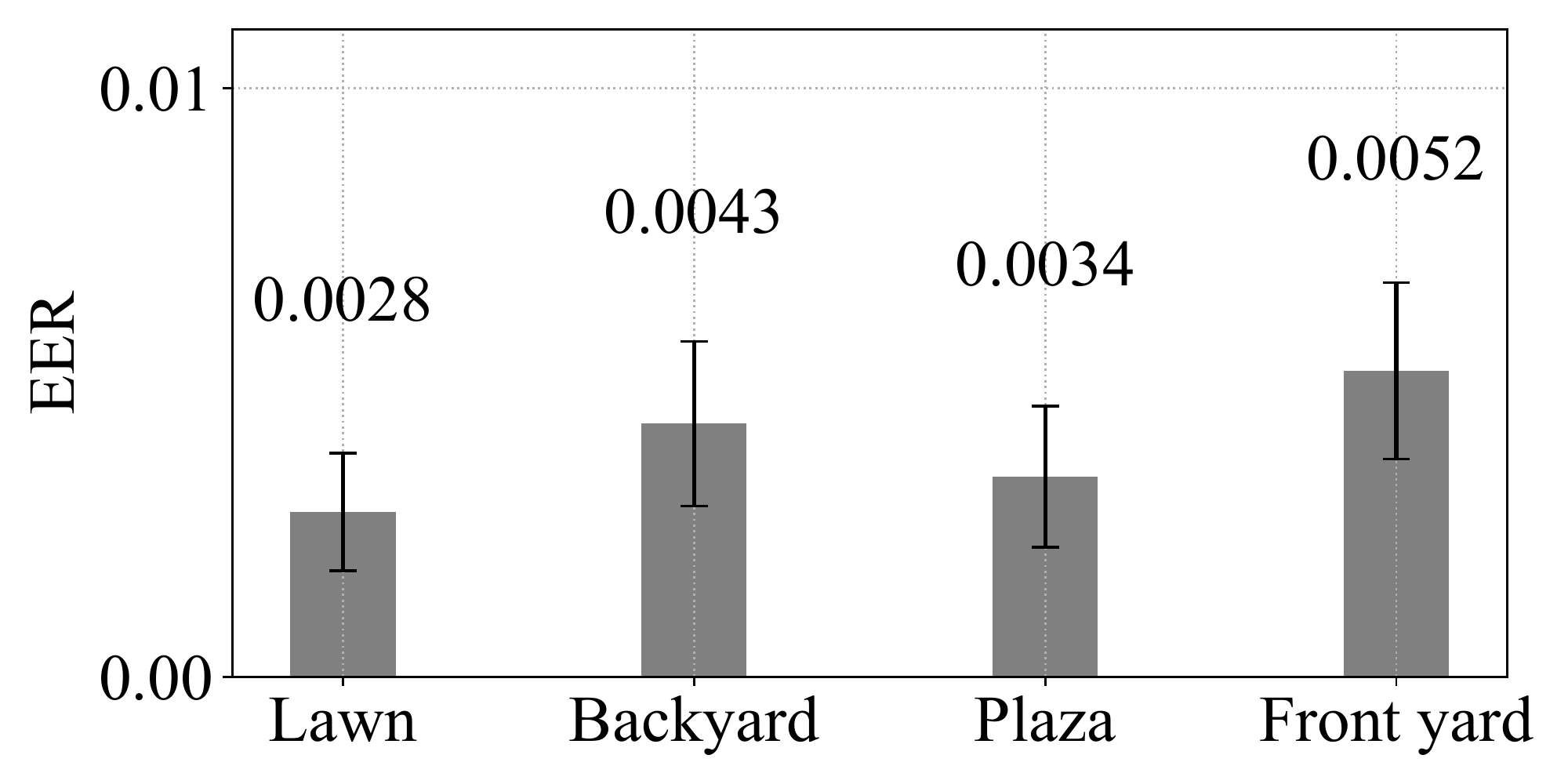}}~ \\
\aaf
\caption{Impact of different parameters and experimental settings. 
}
\label{fig:parameter}
\end{figure*}

\Para{Classifier.}
We train the model with different commonly used classifiers, including SVM, kNN and Random Forest. For SVM, we examine three kernels,  \ie linear, polynomial and radial basis function (RBF) kernels; after grid search, we finally adopt the linear kernel and the optimal hyper-parameters are set as follows, $c$ as 5.0 and $\gamma$ as 0.001.
For $k$NN, we vary the values of $k$ from 1 to 20, and finally set 13 as the optimal parameter.
For Random Forest, we test the model by varying the number of trees from 50 to 200, and adopt the optimal value as 90. 
The results (EER$_{\text{SVM}}$=0.0030, EER$_{\text{RF}}$=0.0032, EER$_{k\text{NN}}$=0.0038) show that SVM has the lowest EER. 
Thus we adopt the linear SVM classifier. 

\para{Duration of audios.}
A longer audio provides richer information but
harms usability as more time is needed for authentication. 
We vary the duration from $1s$ to $5s$ in steps of $1s$.
Initially, EER decreases as the duration increases. When it is $\ge 3s$, EER keeps stable. We thus chose $3s$.

\para{Training dataset size.}
We study how the size of the training dataset impacts the accuracy. We train \shortname by varying the the number of samples, and then test the model with the rest of the samples that are not used for training. Figure~\ref{fig:parameter}~(a) shows that the EER of the classifier converges, given the training dataset has more than 1400 samples. Thus, the 2000 data samples collected in \emph{Dataset I} are sufficient.  

\para{Distance between verifier and drone.}
We evaluate the impact of the drone-phone distance on accuracy. We keep the drone hovering at the height of $4m$ for the 
sake of safety and vary the drone-phone vertical distance from 4m to 8m in steps
of 1m. We collect 2000 samples at each distance. As shown in Figure~\ref{fig:parameter}~(b), the EER rises as the distance  increases. In addition, the EER grows faster than the distance increases. This is because acoustic attenuation follows the inverse square law during its propagation in the air~\cite{wiener1958sound}. 
When the distance is $\le 5m$ meters, the EER grows slowly. 
To balance accuracy and security, we recommend a distance around $5m$. The drones employed in our experiment are compact and inexpensive models. It's worth noting that larger drones often generate louder noises, which in turn enhances the effective authentication distance between service drones and verifiers.

\para{Environmental sounds.}
To evaluate how well our system works 
in different environments, 
we test \shortname in realistic environments compliant with regulations of Federal Aviation Administration (FAA)~\cite{faa_regulation}: (1) \emph{a lawn 15 meters away from a highway}, (2) \emph{a backyard with a party going on}, (3) \emph{a plaza located in a town}, (4) \emph{a front yard with a lawnmower working}. For each environment, we test \shortname in the time period with some of the most complex background noises, \ie when the highway is in a rush hour for the lawn near the highway, in the party with people talking as well as a loudspeaker playing music, when the plaza is in its most popular hours, and the time when the lawnmower is cutting grass. The average sound level for the lawn is about 70 dB, for the backyard 71 dB, for the plaza 65 dB, and for the front yard 73 dB. 
Figure~\ref{fig:parameter}~(c) illustrates the results. There is no significant difference of the performance between different environments, showing that our system is highly robust under various environmental sounds.

\para{Verifier-side devices.}
We study whether \shortname can work well on smartphones of different models and operating systems as well as \textbf{smartwatches}. We select four more mobile devices: 
(1) Nexus 5X, an old Android smartphone released in 2015, (2) iPhone 11, (3) Google Pixel 4, a high-end Android phone released in 2019, and (4) LG W200, a smartwatch. 
No significant difference is observed in the authentication performance between the verifier-side devices. We thus conclude that \shortname works well on different mobile devices, which indicates that \shortname can benefit a broad range of people owning various mobile devices. 

\para{Unseen drones.} 
We adopt the Leave-One-Subject-Out (LOSO) cross validation mechanism to evaluate how \shortname performs on unseen drones. We iteratively choose one drone for testing and use the data of the other drones to train the system. The LOSO mechanism eliminates object bias and evaluates system performance on unseen objects. The average EER over the drones is 0.003 and the standard deviation is 0.0016. The low EER and standard deviation show that \shortname performs well on all of these drones and there is no significant difference in their accuracy. Thus, \shortname can be generally deployed for different drones without training a specific model for each drone, which is preferable for large deployment. 

\subsection{Resilience to Attacks}
\label{Sec:security_analysis}

We aim to not only defeat attacks that are
ignored in prior work~\cite{sounduav,sound-proof} but also reduce false rejections, in order that
both good security and usability are attained. We use Dataset II to study both
the attack success rate under various attacks and robustness to environmental sounds.

\subsubsection{Resilience to Dominant Sound Attacks}
\label{Subsec:dominant_attack}

\begin{figure*} 
\centering
\aaf\aaf
\subfloat[Dominant sound attack]{\includegraphics[scale=0.28]{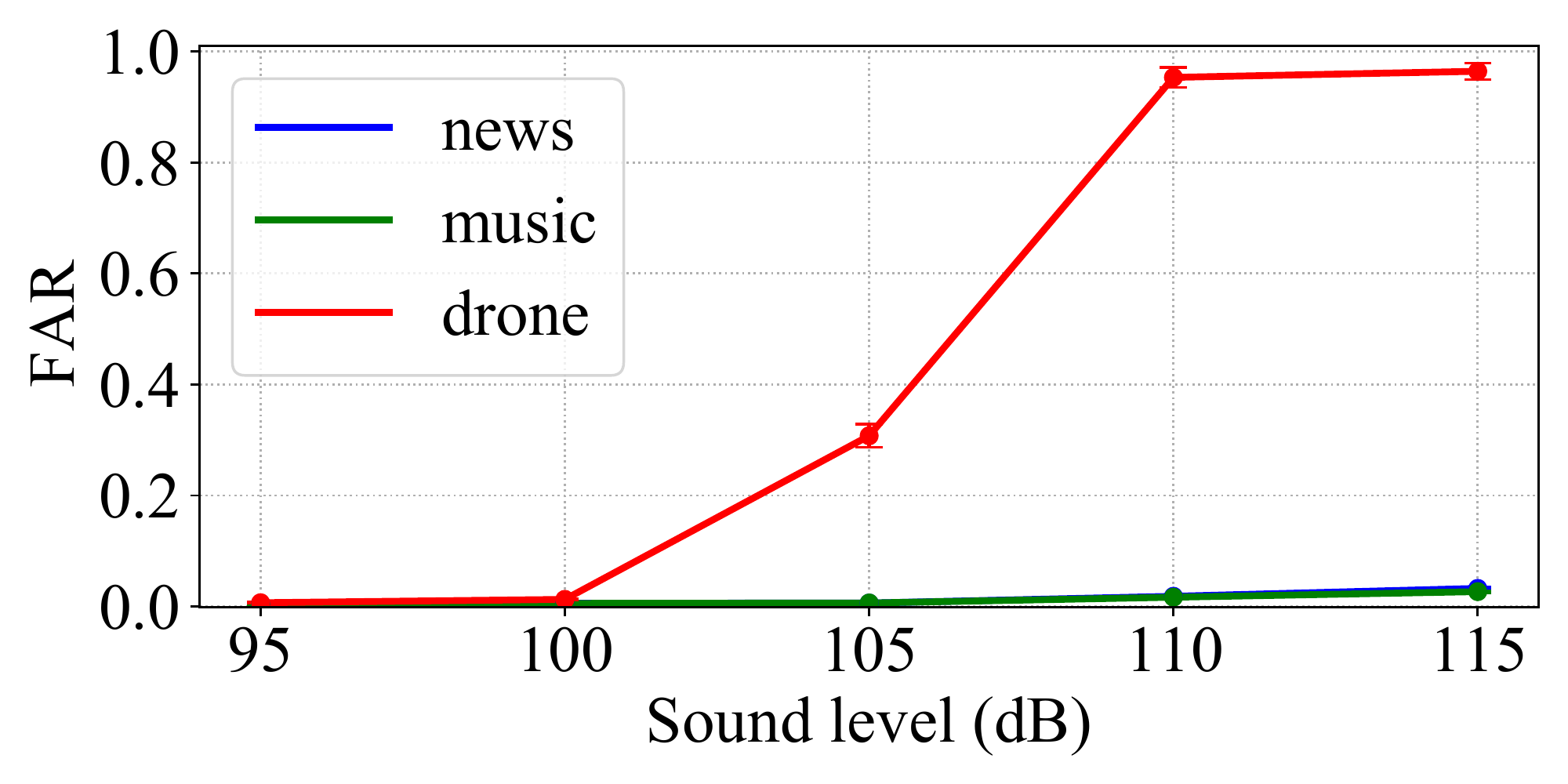}} ~
\subfloat[Verifier-side \audiorelay attack]{\includegraphics[scale=0.28]{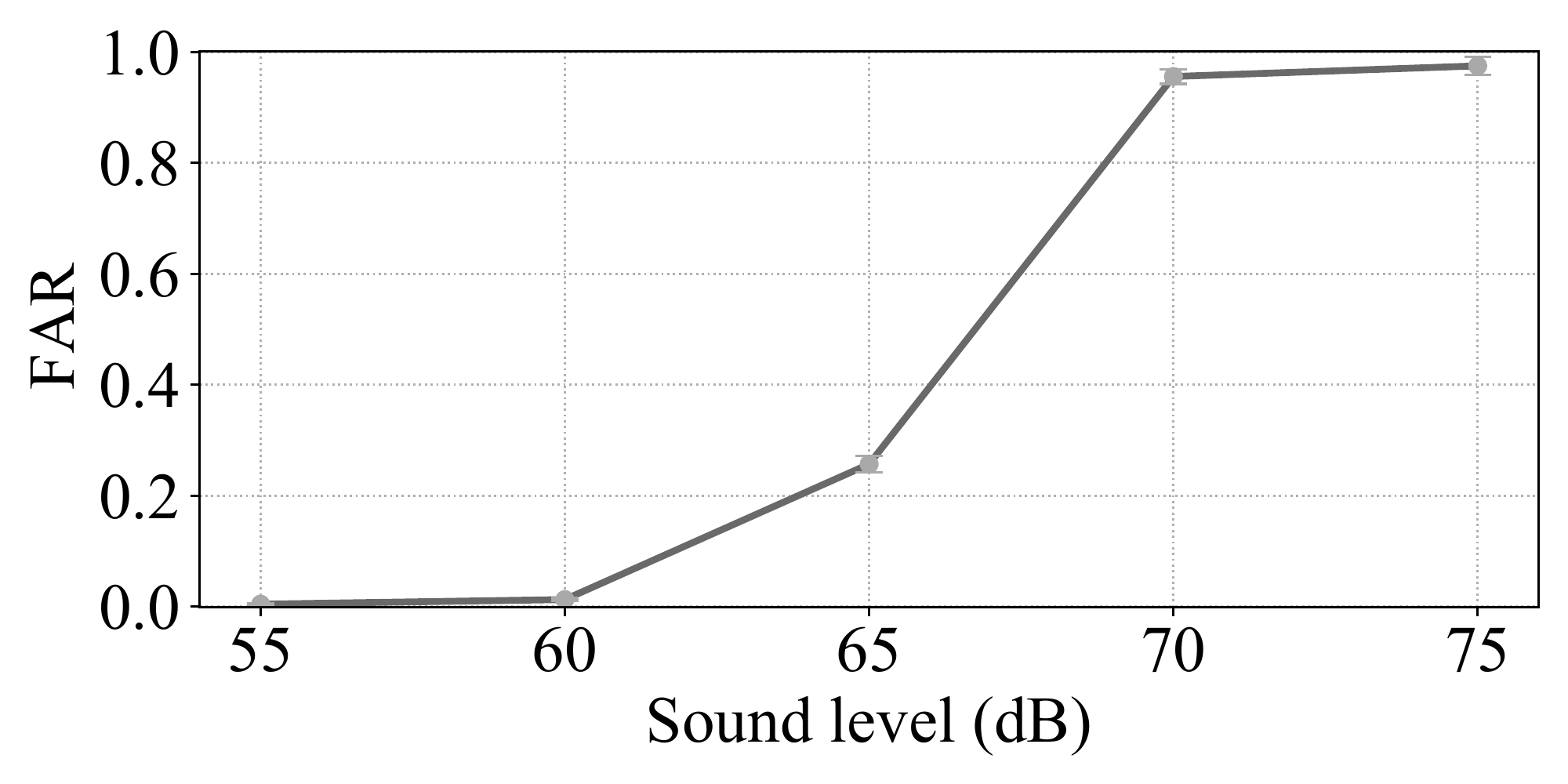}}~
\subfloat[Drone-side \audiorelay attack]{\includegraphics[scale=0.28]{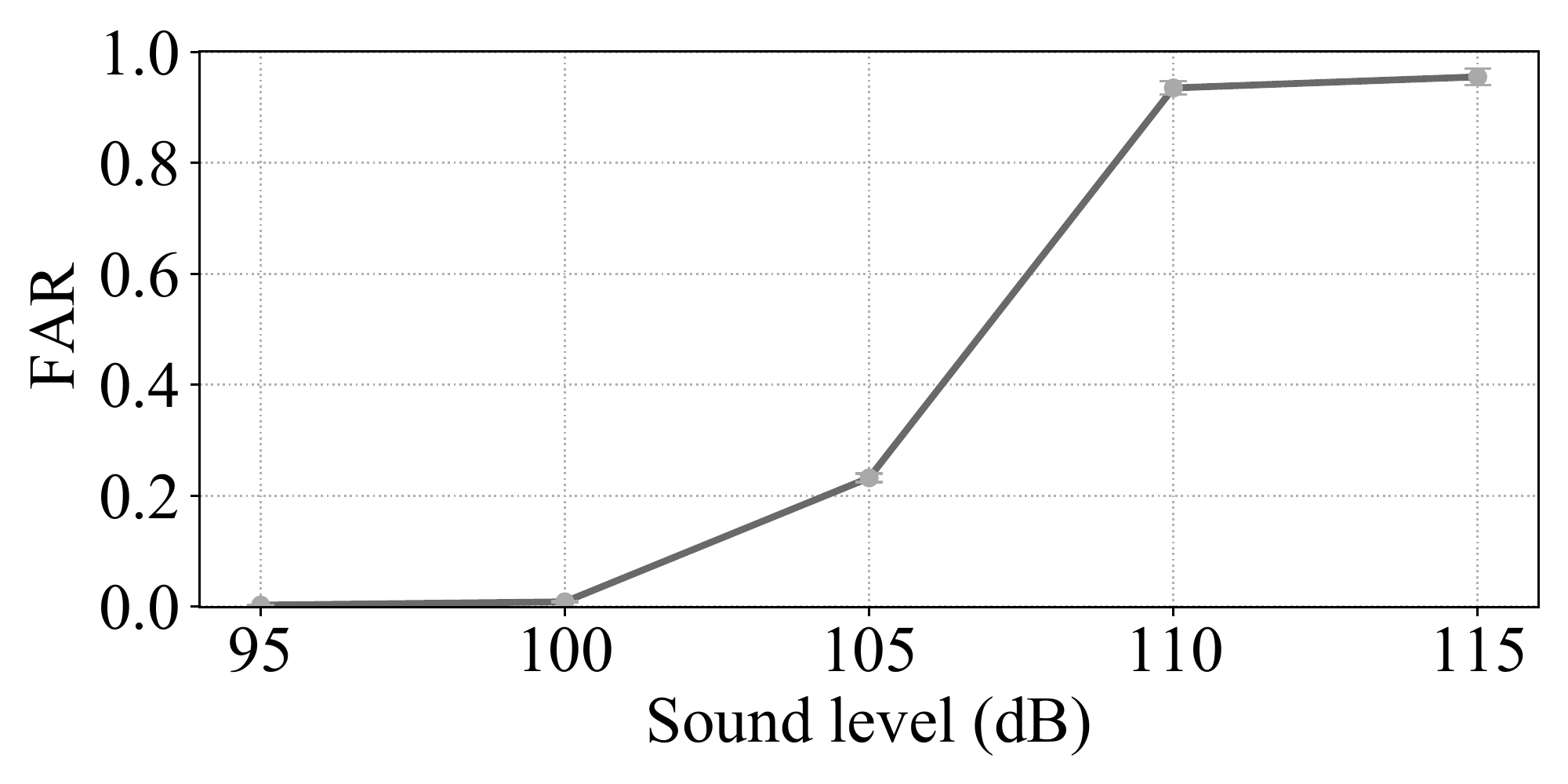}}~
\caption{Attack success rates when \emph{only} audio similarity is used for authentication (i.e., \emph{\textbf{without attack detection}}). This also highlights the
insecurity of prior work~\cite{sound-proof}, which only uses similarity for authentication. 
}
\label{fig:attack_success_rate}
\aaf\af
\end{figure*}

\Para{Attack success rate without attack detection.} 
Figure~\ref{fig:attack_success_rate}(a) shows the attacker's success rate when launching dominant sound attacks under different malicious sound levels from 95 to 115 dB at steps of 5 dB. The attacker has to replay  recorded drone noises, as the malicious sounds, at a very high level ($>$ 100 dB) to effectively fool audio similarity comparison of \shortname. The reason is that a high malicious sound level is required to
\emph{distort} sounds recorded by the victim drone.\footnote{One may propose to not play any malicious sounds on the drone side, but simply record the victim drone noises and replay live on the victim smartphone side. Note that it is \emph{verifier-side audio relay attacks}.} Keep in mind that the drone-side microphone  is closely attached to the drone, so there is little attenuation when it records drone noises. The victim drone sound level measured by the sound level meter attached to DJI Mavic Mini is $99.3 \pm 1.8$ dB. Thus, the attacker needs to play very loud malicious sounds to fool \shortname. 
(If music or news, instead of recorded drone noises, is used as malicious sounds,
the attack success rates keep low. This is because music and news have less energy in the frequencies used for similarity determination. Thus, a smart attacker should
use drone noises as the malicious sounds for launching dominant sound attacks.)

\para{Effectiveness of sound level checking.}
We then investigate whether we can exploit the observation above to distinguish
attacks from environmental sounds. According to the information provided by CDC~\cite{cdc_hearing}, a human may feel very annoyed when the sound level is between 80 dB and 85 dB, and sounds of 95 dB are considered damaging to hearing.
On the other hand, an attacker needs to impose ``environmental sounds'' $>$ 100 dB to 
fool \shortname. Thus, by checking
whether the environmental sounds are louder than 100 dB, we can detect all the dominant sound attacks that are \emph{otherwise} successful, and do not cause false
rejections because of environmental sounds (unless they are extremely loud).

We have the following observations. First, the experiment above uses 
the DJI Mavic Mini drone, which generates the smallest noises. According to our experiments with other drones, the malicious
sounds have to be even louder in order to distort the drone noises and fool the audio similarity comparison of \shortname, which makes it even easier for
the sound level meter to
distinguish the even louder malicious sounds from everyday environmental sounds. 

Second, one may propose to further enhance the method by adding
content detection.
When the content is detected as music or news, it can allow environmental sounds up to 110 dB
without worrying about successful attacks (Figure~\ref{fig:attack_success_rate}(a)). 
However, most benign environmental sounds are 
less than 95 dB. We thus consider this enhancement non-critical.

Finally, dominant sound attacks need an attacker to impose malicious sounds to both
the victim drone and the victim smartphone. Since the microphone of the victim smartphone is
multiple meters away from the loud hovering drone, its recording is easier to distort. 
We thus focus on the drone side to design the attack detection method.

\subsubsection{Resilience to Verifier-side Audio Relay Attacks}
\label{Subsec:user_side_replay_attack}

\begin{figure*}
\centering
\aaf
\subfloat[EER \emph{vs.} loudspeakers]{\includegraphics[scale=0.28]{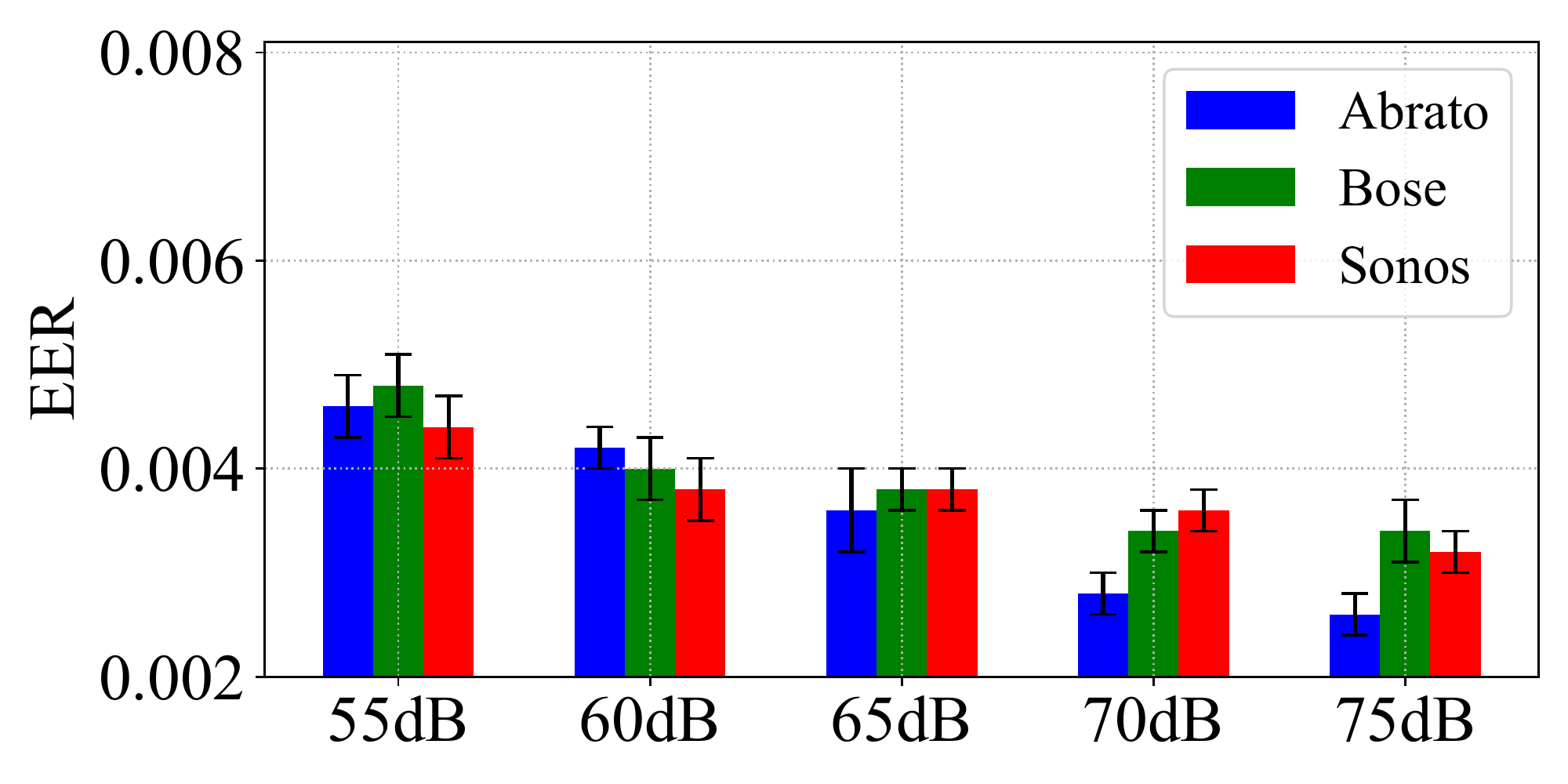}}~
\subfloat[EER \emph{vs.} verifier-side devices]{\includegraphics[scale=0.28]{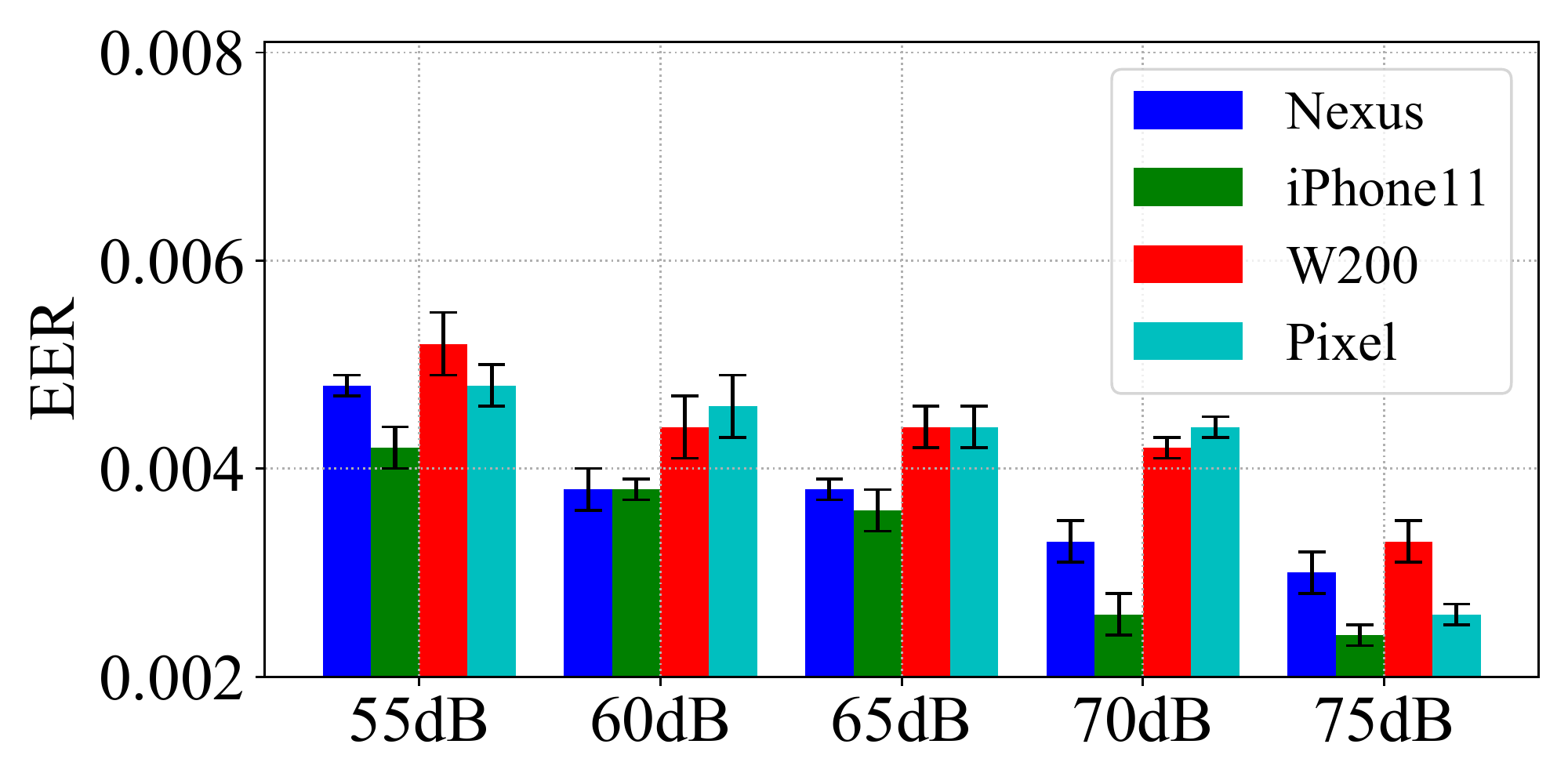}}~
\subfloat[EER of LSP content detection]{\includegraphics[scale=0.28]{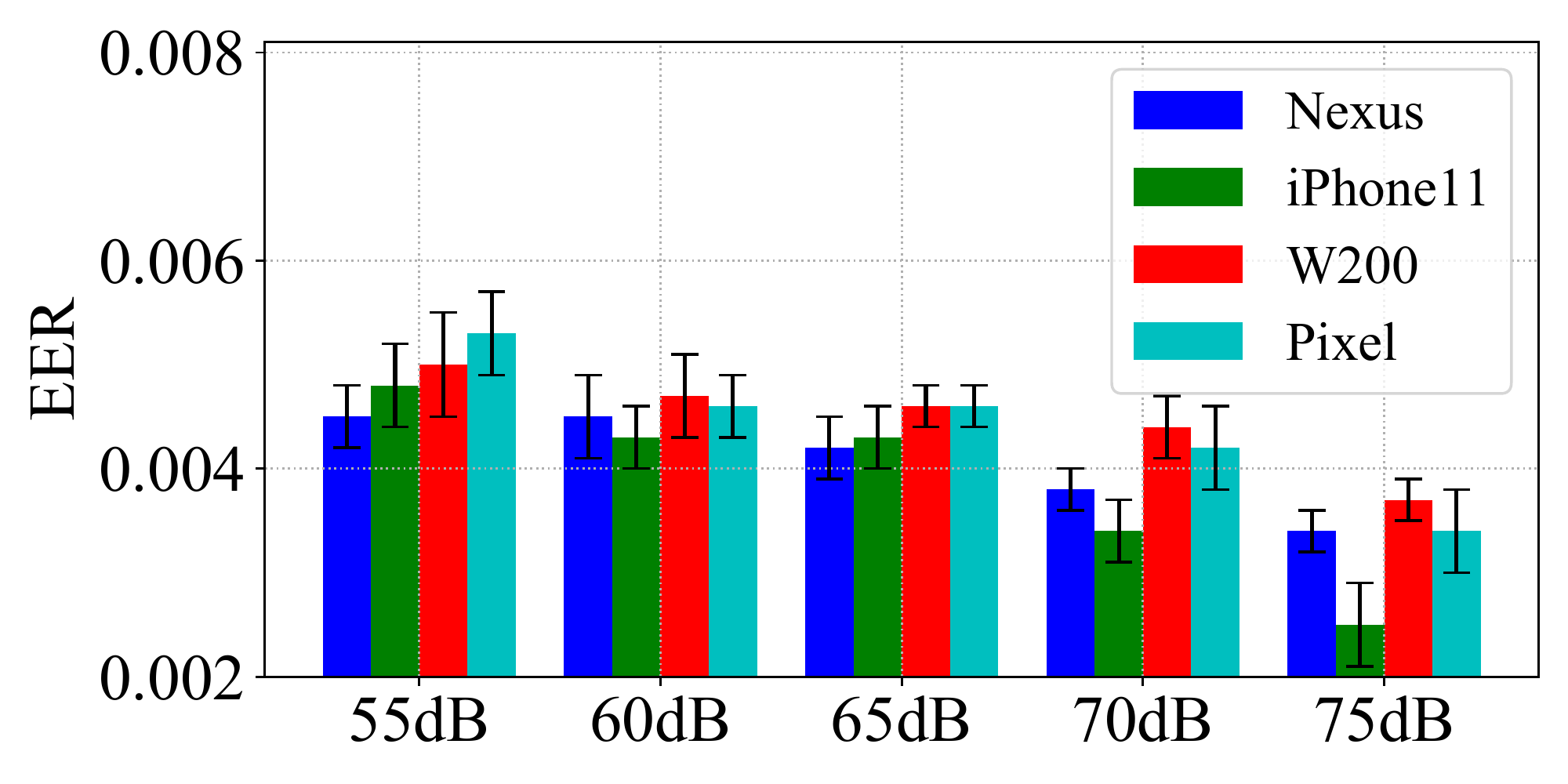}} ~
\caption{Effectiveness of drone sound liveness checking and loudspeaker content detection. 
The EER
keeps very low. (When the malicious sound level is low, e.g., 60 dB, the EER
is slightly higher, which is because  the loudspeaker's characteristics are less intensive with lower sounds and thus more difficult to detect. But a low malicious
sound level cannot fool our audio similarity and is thus not threatening; see Figure~\ref{fig:attack_success_rate} (b).)
}
\label{fig:speaker_detection}
\aaf\aaf
\end{figure*}

\Para{Attack success rate without attack detection.} Figure~\ref{fig:attack_success_rate}(b) shows the success rate of verifier-side \audiorelay attacks when \emph{only} similarity comparison is conducted for authentication, i.e., \emph{without attack detection}. By raising the volume of the loudspeaker, the attacker increases his success rate quickly. Once the recorded sound reaches to a certain level and dominates the drone's sound, the attacker's success rate becomes very high. For example, when the malicious sounds recorded on the verifier side is 70dB, the attack success rate is 0.955. This alarming result illustrates the insecurity of prior work like SoundProof~\cite{sound-proof} that does not tackle such attacks.

\para{Effectiveness of drone sound liveness detection.} 
We extract features described in Section~\ref{subsubsec:speaker_detection} and use SVM to train a model based on the data collected above. To evaluate the \emph{generalizability} of the drone sound liveness detection algorithm, we use the strict \emph{Leave-One-Subject-Out} (LOSO) method: we train the model on the data collected with two loudspeakers and test using the data collected with the third unseen loudspeaker, and repeat the process to evaluate all the loudspeakers. The result is shown in Figure~\ref{fig:speaker_detection}(a). The low EER indicates that the loudspeaker detection algorithm has a high generalizability, which means the model can be used to distinguish \emph{unseen} loudspeakers. The higher the malicious sound level is, the lower the EER, as the loudspeaker's characteristics get more intense.

We also use LOSO to evaluate the \emph{generalizability} of the drone sound liveness detection algorithm over different smartphones. For each unseen smartphone, we train a model with data collected by the other smartphones and test the model with data of the unseen phone. Figure~\ref{fig:speaker_detection}(b) shows the results. The EER is low and stable over all the sound levels for all the smartphones, which means the model can be used by different smartphones under various sound levels.

\para{Effectiveness of loudspeaker content detection.} 
We play various music and news via a loudspeaker and record the sounds while having a drone fly nearby. We change the sound level from 55 dB to 75 dB in steps of 5 dB and for each sound level, we collect 1000 pieces of data for playing music and news each; each data piece is denoted as positive. The negative data is collected by having the loudspeaker play recorded sounds produced by a drone. We 
also use the strict LOSO method to train and
test the SVM model, in order to evaluate the effectiveness when an unseen smartphone is used. Figure~\ref{fig:speaker_detection}(c) shows that our system can distinguish music and news from replayed drone noises at a high accuracy, even if an unseen smartphone is used. 

\subsubsection{Resilience to Drone-side Audio Relay Attacks} 
\label{Subsec:drone_side_replay_attack}

Figure~\ref{fig:attack_success_rate}(c) shows the attack success rate when there is
no attack detection. Like the dominant sound attack, when the malicious sound
level is larger than 100 dB, the attack success rate grows quickly. Given the
malicious sounds are loud, the sound level meter can easily distinguish
malicious sounds from everyday environmental sounds.

\subsubsection{Summary of Resilience to Attacks} 
The thorough evaluation demonstrates
the following. (1) The attack
success rate would be very high \emph{without} countermeasures (see
Figure~\ref{fig:attack_success_rate}), while prior works that make
use of sounds for authentication, such as Sound-Proof~\cite{sound-proof},
failed to handle these attacks. (2) \shortname
is highly resilient to various attacks. (3) \shortname
is robust under various environmental sounds.

\subsection{{Authentication Time}}
\label{Subsec:efficiency}

\begin{table}
    \centering
    \caption{Authentication time.}
    \aaf\af
    \resizebox{.85\columnwidth}{!}{%
    \begin{tabular}{c|c|c}
        \hline
        \textbf{Part} & \textbf{Mean (ms)} & \textbf{Std. Dev. (ms)} \\ \hline
        Audio Recording & 3,000 & - \\\hline
        Data Transmission & 964 & 117 \\ \hline
        Decision Making & 749 & 76 \\ \hline
  \end{tabular}
  }
\label{tab:efficiency}
\aaf\af
\end{table}

We evaluate the average authentication time needed for \shortname, measured from the audio recording start time to a decision made. It mainly contains three parts: (1) time for audio recordings; (2) time for data transmission; and (3) time for  decision making. Time for each part is shown in Table~\ref{tab:efficiency}. The \emph{\textbf{total time}} for authentication  is $4.71 \pm 0.19$s on average. Thus, \shortname works quickly.

\section{Usability Study}
\label{sec:user_study}
The objective of our usability study is to determine user acceptance of \shortname in comparison to other authentication methods. Specifically, we are comparing \shortname with facial recognition and fingerprint scanning-based authentication, which are increasingly being used in mobile devices~\cite{yang2020evaluating, fernandez2016small}. Although facial recognition and fingerprint scanning are
not used for drone authentication, they are widely used authentication
methods in other scenarios.  
The comparison focuses on usability rather than security. We did not
inform the participants that \shortname is a mutual authentication mechanism while the other two can only be used to authenticate one party.

\begin{figure*} 
\centering
\subfloat{\includegraphics[scale=0.37]{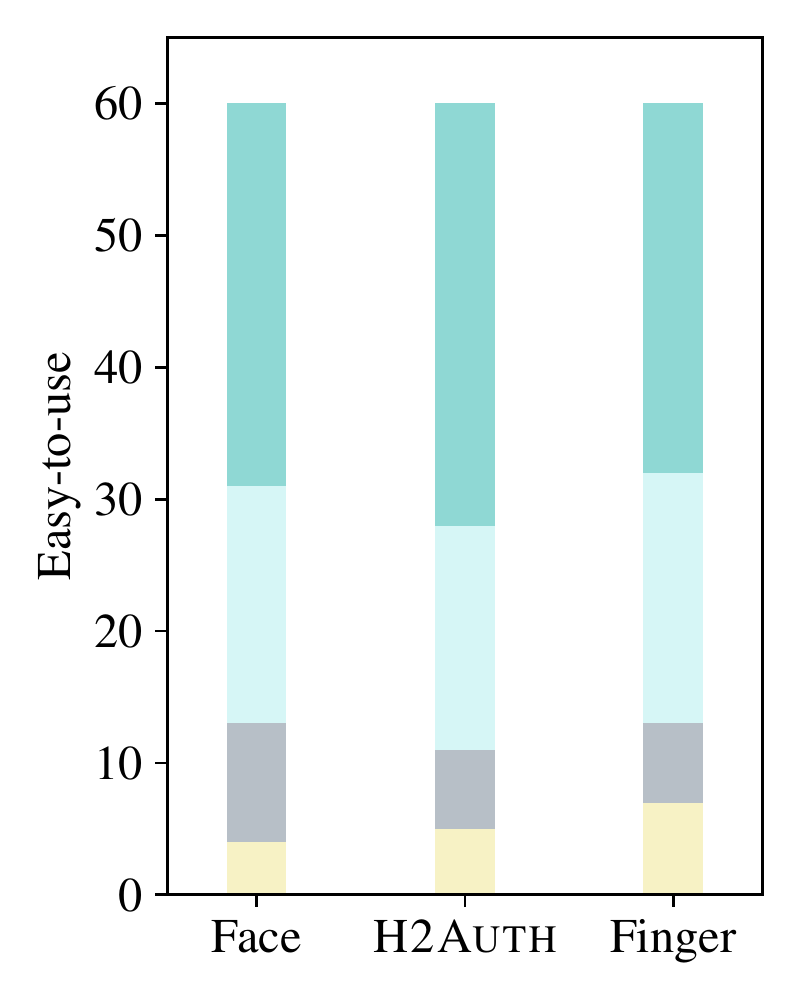}}
\subfloat{\includegraphics[scale=0.37]{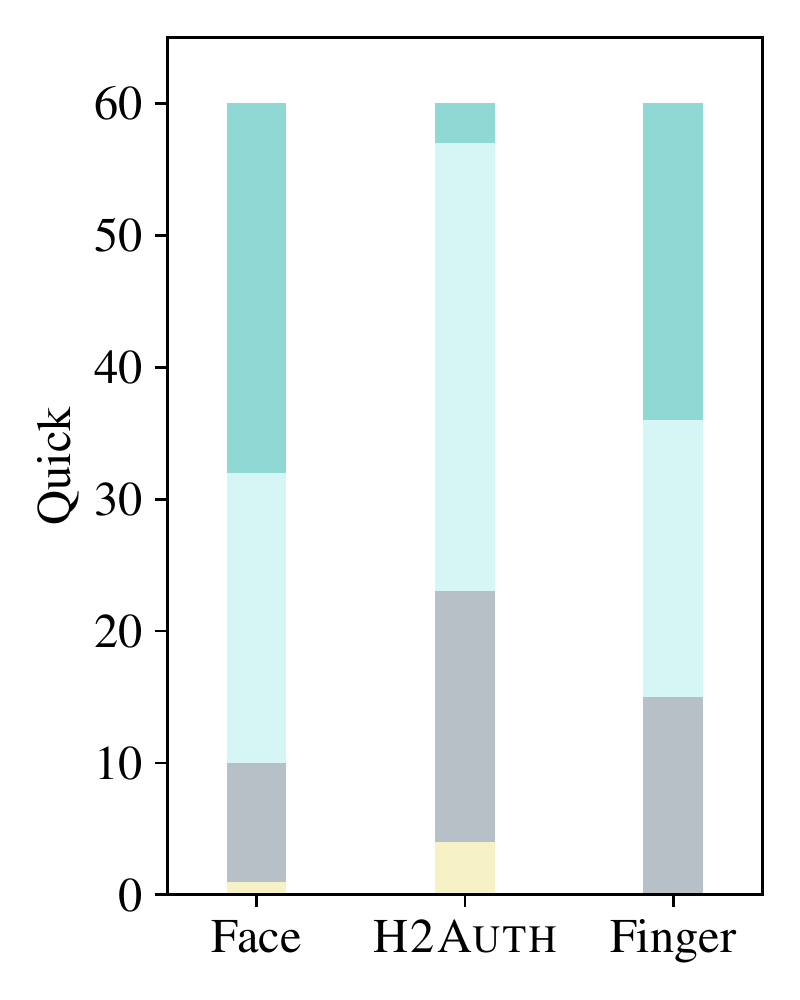}}
\subfloat{\includegraphics[scale=0.37]{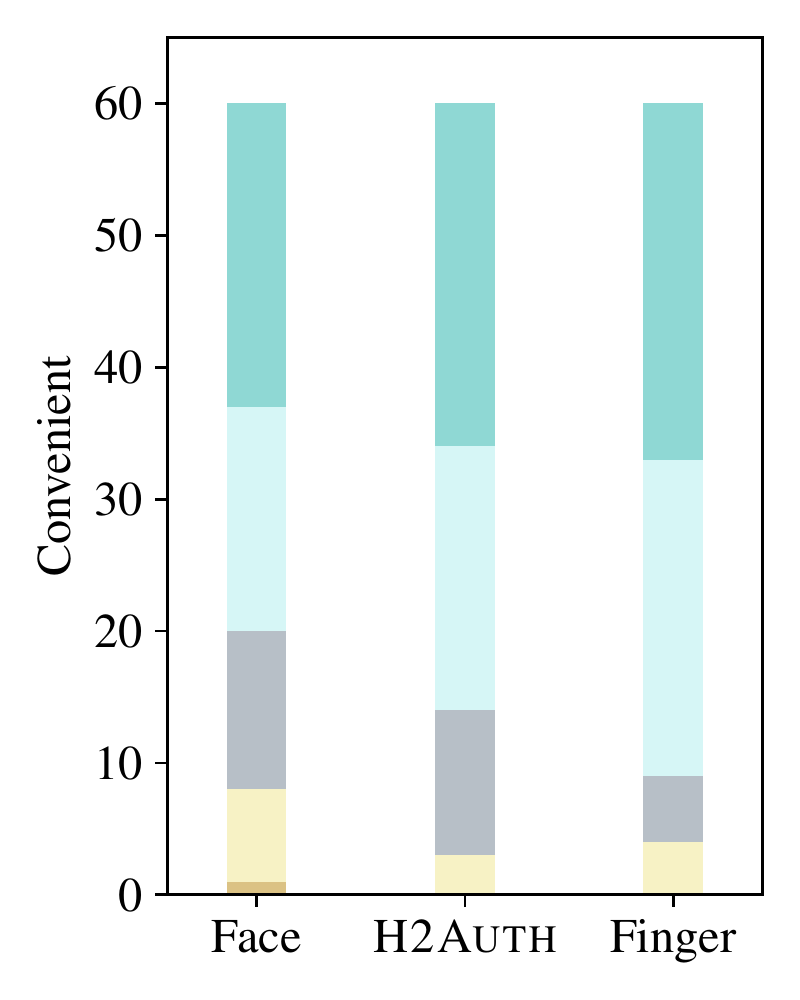}}
\subfloat{\includegraphics[scale=0.37]{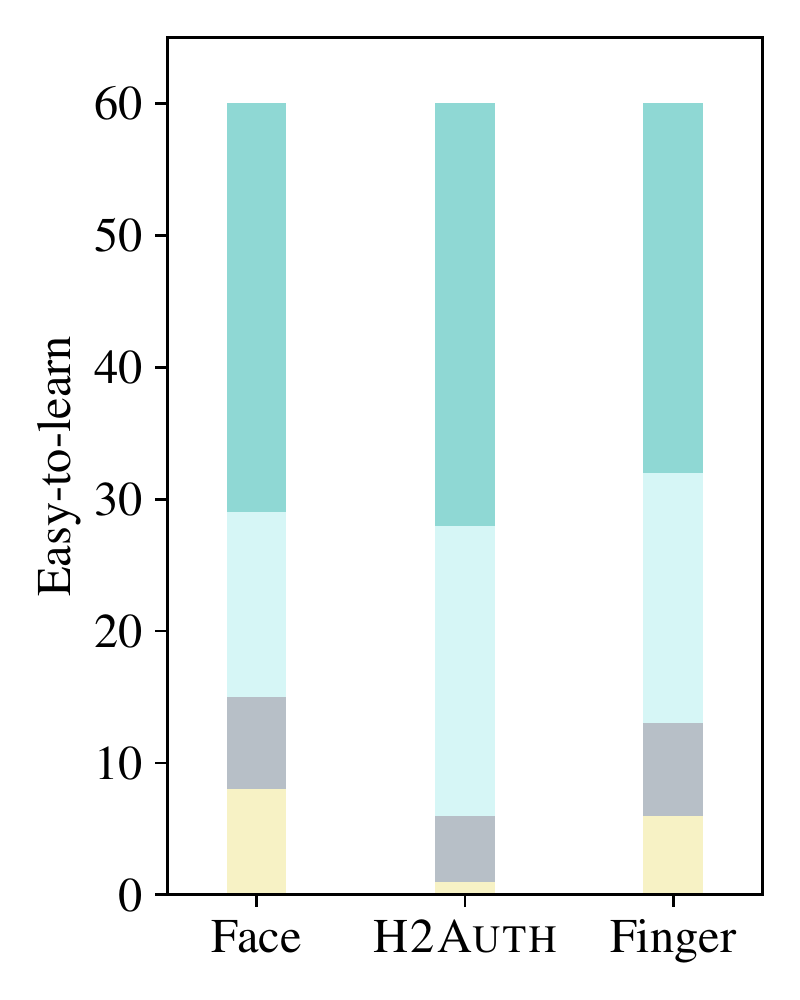}}
\subfloat{\includegraphics[scale=0.37]{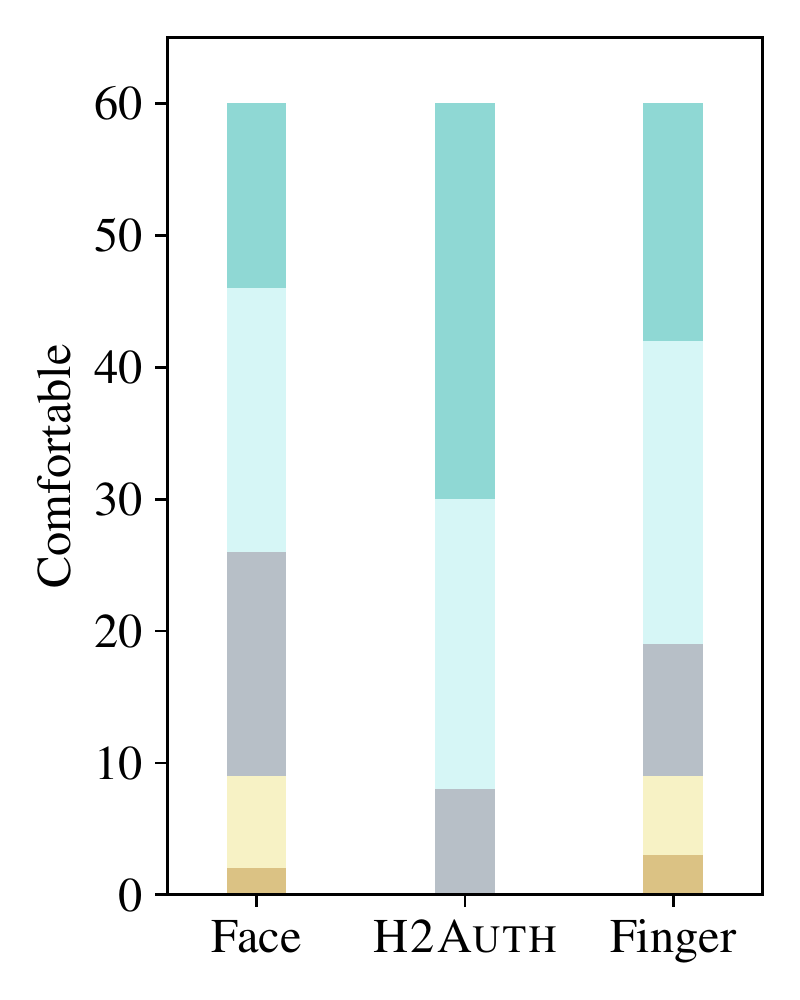}}
\subfloat{\includegraphics[scale=0.37]{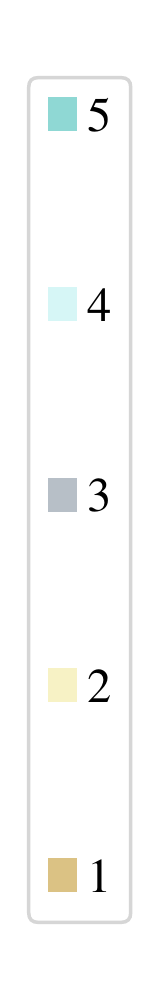}}
\aaf
\caption{Distribution of the answers to the usability study  questions.}
\label{fig:usability}
\aaf\aaf
\end{figure*}

\aaf
\subsection{Recruitment and Design}
We recruited 60 participants using a snowball sampling method, the majority of whom were not from our department and did not have a computer security background. To prevent social desirability bias, we did not disclose that \shortname was developed by us. Instead, participants were instructed to evaluate the usability of various authentication methods. The usability study was conducted with an IRB approval, and each participant was required to sign a consent form.

To introduce the three authentication methods to the participants, we played a pre-recorded video. For the facial recognition method, we provided an iPhone 11 equipped with FaceID, while for the fingerprint scanning method, we used an iPhone 6s plus with TouchID to scan and profile the participants' fingerprints. Participants were then instructed to set up their FaceID and TouchID. Following this, each participant performed five authentication attempts for each method to become familiar with them. These initial attempts were excluded from further analysis. Next, each subject performed three additional attempts using each authentication method, with the order of methods used randomized.

At the conclusion of the study, each participant completed a questionnaire evaluating the three authentication methods by answering five questions, adapted from the widely-used SUS~\cite{brooke1996sus}. The five questions were as follows: \emph{(1) I found the authentication method easy to use; (2) I was satisfied with the amount of time it took to complete the authentication; (3) I found the authentication method convenient; (4) I believe it is easy to learn the authentication method; and (5) I felt comfortable using the authentication method.} Scores for each question ranged from 1 to 5, with 1 indicating strong disagreement and 5 indicating strong agreement. Higher scores indicated better usability.

\aaf
\subsection{Usability Study Results}
\Para{Demographics.} Out of the 60 participants, 52\% were female and 48\% were male. Among them, 5\% were between 15 and 20 years old, 55\% were between 21 and 30 years old, 28\% were between 31 and 40 years old, and the remaining 12\% were above 40 years old. All participants used the fingerprint scanning-based authentication method, with 78\% of them having used TouchID before. Additionally, 97\% of the participants used the facial recognition-based authentication method, with 43\% having used FaceID before.

\para{Perceived usability.} The questionnaire included five questions that evaluated usability based on the following aspects: ease-of-use, speed, convenience, ease-of-learning, and comfort. The distribution of scores for each authentication method is shown in Figure~\ref{fig:usability}. The results indicate that users found \shortname to be easy-to-use, convenient, easy-to-learn, and comfortable. However, it was not perceived as being as quick as the other two methods.

The total scores for facial-recognition-based method, \shortname, and fingerprint-scanning-based method are $20.13 \pm 2.28$, $20.80 \pm 1.90$, $20.45 \pm 2.47$, respectively. 
The scores show that \shortname achieves slightly better perceived usability than FaceID and TouchID methods. The main difference between \shortname and the other two methods lies in how comfortable the user feels. The other two methods require users 
to enroll in the system by profiling their face or fingerprint while \shortname does not use any biometrics. This is confirmed by some comments left by the participants, saying that they did not like public devices collecting 
their personal information. 

The average authentication time for using \shortname, FaceID,  and TouchID approach is $4.7 \pm 0.17$s, $1.05 \pm 0.46$s, and $0.95 \pm 0.53$s, respectively. \shortname requires 
more time for authentication than the other two methods. However, the authentication time of 4.7 seconds is still acceptable for a drone service, as 37 of the 60 participants thought it was quick to use \shortname for authentication.
\af
\section{Related Work}
\label{Sec:Related_Work}
Authentication based on information correlation is a promising direction and has inspired many great works~\cite{touch_to_access, mare2018saw, zebra, li2019touch, ruiz2018idrone,wu2022g2auth,sharp2022authentication,wang2016touch, 2014CCS_CZP, shakewell, ccs20, H2H_CCS14, han2018you, wu2022use,yang2023wave}. 
Prior state of the art work on drone authentication, G2Auth~\cite{wu2022g2auth} and  Smile2Auth~\cite{sharp2022authentication}, follows this direction. G2Auth~\cite{wu2022g2auth} has a user hold a smartphone and wave her hand; then it
compares the smartphone's IMU data 
and the drone's video for authentication. Smile2Auth~\cite{sharp2022authentication} has a user change facial expressions and compares the face embeddings collected by the user's smartphone and the drone.
Both G2Auth and Smile2Auth require 
humans (waving or smile), while \shortname can be used for non-human scenarios, such
as a dock,  smart doorbell, garage, or robot (detailed at the end of Section~\ref{sec:detailed_design}). Thus, \shortname has
prominent advantages in  usability, privacy and generalizability. 

Compared to prior work, SoundUAV~\cite{sounduav} and Acoustic Fingerprint~\cite{daaf}, that also makes use of drone noises for authentication, \textbf{our work does not rely on any drone noise fingerprints}.
Thus, the usability from the perspective of drone service companies is much
improved.  Plus, over time because of motor and blade wear, a drone may need to
be fingerprinted once and again. 
Moreover, \shortname tackles various audio attacks, which are not considered in prior work.
Finally, \shortname keeps robust to various environmental noises, which are not 
examined in prior work.

\section{Discussion}
The arms race between attackers and defenders never ends.
For example,  attackers may manipulate recorded drone noises (e.g., through adversarial deep learning~\cite{abdullah2021sok}) and replay them 
to fool  sound liveness checking. First, the method of sound liveness
checking keeps evolving and becomes increasingly accurate~\cite{zhang2016voicelive,meng2022your}, which
can benefit our authentication system.
Second, during our authentication, a drone randomly and slightly
moves. Thus, it is impossible for a manipulated record prepared in advance
to match the live noises of a randomly moving drone. 
Third, 
the noise recording by a drone should match
the random and slight movements of the drone. For example, when
a drone speeds up, the noise level goes high and the essential frequency (Section~\ref{subsec:preliminary_study}) increases. A drone has the ground
truth about the movements, which can be leveraged
to further enhance our attack detection. We leave this as future work.

\shortname requires the verifier to be equipped with a microphone for capturing the sounds of the drone. While this prerequisite may pose deployment limitations in certain real-world settings, it is worth noting that many contemporary devices, including smartphones, smartwatches, and increasingly common smart home devices like doorbells and cameras, come with built-in microphones. As the prevalence of such smart devices continues to grow, \shortname stands to find application across a wider range of scenarios.

\aaf
\section{Conclusion}
We presented \shortname,  a highly secure and usable mutual authentication solution for drone services that does \emph{not} rely on any drone noise fingerprints. To cope with the sound disparity between drones and verifier devices and tolerate various environmental noises, we devise a novel audio comparison method that exploits the characteristics of drone noises, attaining high accuracy and robustness. 
This is the first work that defeats various attacks against a sound-based mutual authentication system, which distinguishes our system from prior work. 
\shortname can be used in non-human scenarios,
leading to unique advantages in usability and user privacy over prior state of the art.
The comprehensive evaluation shows that it is highly accurate, resilient to attacks, and
robust under environmental sounds.
We envision \shortname can greatly foster secure drone services.

\bibliographystyle{IEEEtran}
\bibliography{reference.bib} 

\vspace{-22pt}

\begin{IEEEbiography}[{\includegraphics[width=1in,height=1.25in,clip,keepaspectratio]{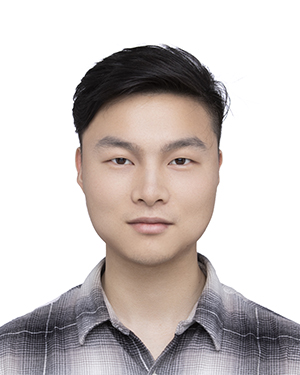}}]{Chuxiong Wu}
(Student Member, IEEE) received the BE degree in Electronic Information Engineering from Beihang University and the MS degree in Computer Science from the University of South Carolina. He is currently working towards the PhD degree in the Department of Computer Science with George Mason University. His research interest focuses on Security in Cyber-Physical Systems.
\end{IEEEbiography}

\vspace{-22pt}

\begin{IEEEbiography}[{\includegraphics[width=1in,height=1.25in,clip,keepaspectratio]{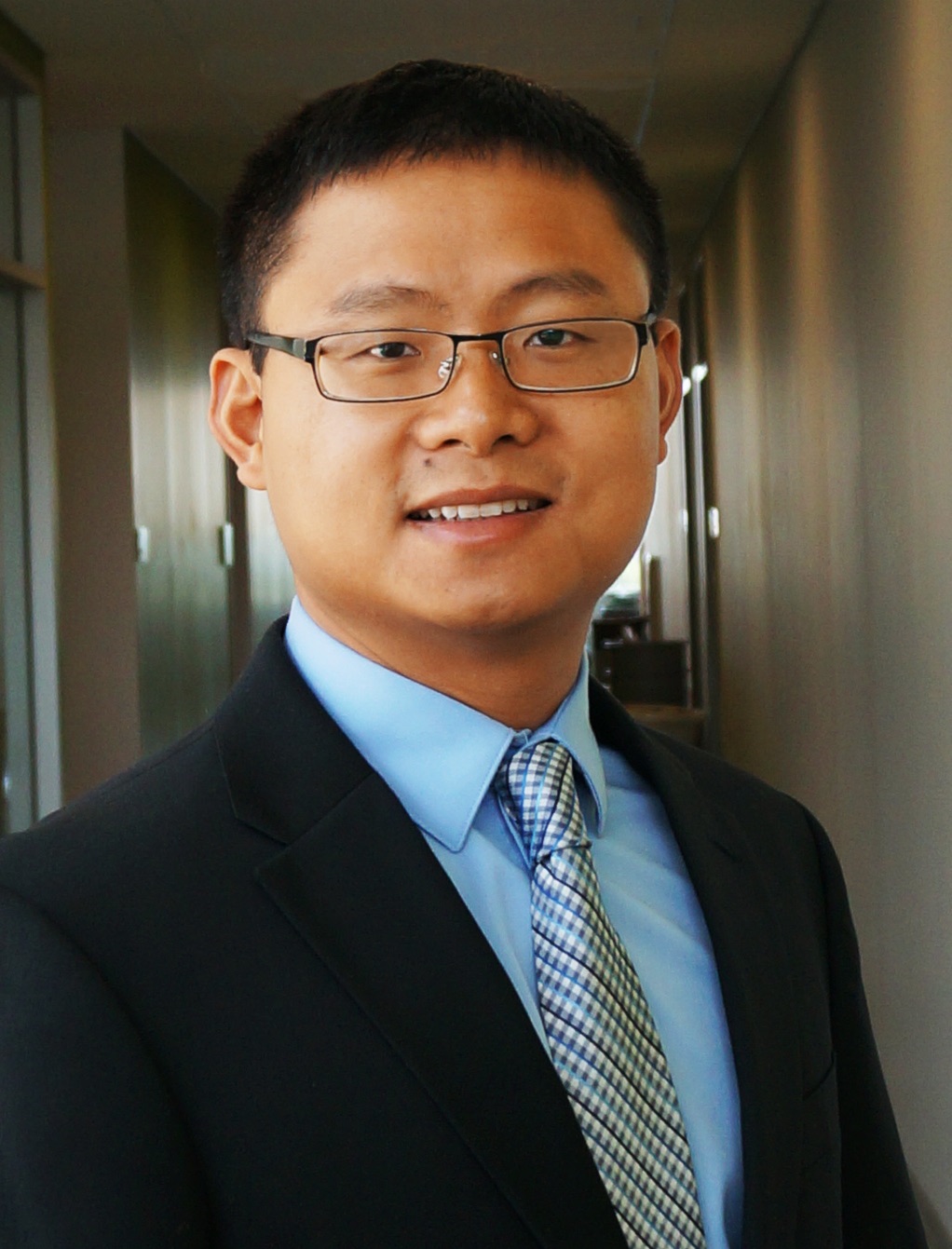}}]{Qiang Zeng}
received the bachelor’s and master’s degrees from Beihang University, and PhD degree from Penn State University. He is an Associate Professor in the Department of Computer Science with George Mason University. He is the recipient of an NSF CAREER Award. His main research interest is Computer Systems Security, with a focus on Cyber-Physical Systems, Internet of Things, and Mobile Computing. He also works on Adversarial Machine Learning.
\end{IEEEbiography}

\end{document}